\documentstyle[12pt,aasms4]{article}
\begin{document}

\title{The X-ray spectra and spectral variability of intermediate type
Seyfert galaxies: $ASCA$ observations of {NGC~4388} and {ESO~103--G35}}

\author{Karl Forster $^{1,2}$, Karen M. Leighly $^{1}$, 
and Laura E. Kay $^{3}$}
\affil{$^{1}$ Dept. of Astronomy, Columbia University,
550 West 120th Street, New York, NY 10027}
\affil{$^{2}$ Current address: Harvard-Smithsonian Center for Astrophysics,
60 Garden Street, Cambridge, MA 02138}
\affil{$^{3}$ Barnard College, Dept. of Physics \& Astronomy, 3009 Broadway, 
New York, NY 10027}
\affil{Electronic mail: kforster@head-cfa.harvard.edu}
\vskip 1cm
\affil{Submitted to the Astrophysical Journal}
\received{29th January 1999}
\accepted{5th March 1999}

\begin{abstract}

The X-ray spectra of two intermediate type Seyfert galaxies are
investigated using $ASCA$ observations separated by more than a
year. Both {NGC~4388} and {ESO~103--G35} exhibit strong, narrow Fe~K$\alpha$
line emission and absorption by cold neutral gas with a column density
$\sim 10^{23}$ cm$^{-2}$, characteristic of the X-ray spectra of type
2 Seyfert galaxies. The power law continuum flux has changed by a
factor of 2 over a time-scale of $\sim 2$ years for both objects,
declining in the case of {NGC~4388} and rising in {ESO~103--G35}.  

No variation was observed in the equivalent width of the Fe~K$\alpha$
line in the spectra of {NGC~4388}, implying that the line flux
declined with the continuum. We find that the strength of the line
cannot be accounted for by fluorescence in line-of-sight material with
the measured column density unless a `leaky-absorber' model of the type 
favored for {IRAS~04575--7537} is employed. 

The equivalent width of the Fe~K$\alpha$ emission line is seen to 
decrease between the observations of {ESO~103--G35} while the continuum 
flux increased.  The 1996 observation of {ESO~103--G35} can also be fitted 
with an absorption edge at 7.4 $\pm$ 0.2 keV due to partially ionized iron, 
and when an ionized absorber model is fitted to the data it is found that the
equivalent column of neutral hydrogen rises to $3.5 \times 10^{23}$
cm$^{-2}$. The Fe K$\alpha$ line flux can be accounted by
fluorescence in this material alone and this model is also a good
representation of the 1988 and 1991 $Ginga$ observations. There is
then no requirement for a reflection component in the $ASCA$ spectra
of ESO~103--G35 or NGC~4388.

\end{abstract}


\section{Introduction}

The taxonomy of Seyfert galaxies has expanded from the straightforward
types 1 and 2 classification to include various intermediate (1.2,
1.5, 1.8, and 1.9) types as well as type~3 or LINERS (Heckman 1980;
Filippenko 1996). This system is primarily based on the strength of
the broad permitted emission lines of the Balmer series which are
broad compared to the width of the emission lines from forbidden
atomic transitions such as {[\ion{O}{3}]$\lambda 5007$}.  There are
now reasons to believe that what we are seeing is the result of a
viewing angle dependence of optically thick material that obscures the
Broad Emission line Region (BLR) formed in a toroidal distribution. 
The observation of broad Balmer lines in the
polarized optical spectra, broad infrared emission lines of the
Paschen series, and the large column densities measured in the X-ray
spectra of some type 2 Seyferts lend support to what has become a
unified model of Seyfert galaxies.

X-ray observations give a direct measure of the column density of 
line-of-sight material absorbing the continuum emission of the AGN.  
Comparison between the variations of the hard X-ray continuum and the 
Fe K$\alpha$ fluorescence emission line, produced by the transmission 
of X-rays through the absorbing material, allow an estimate of the 
location of the torus to be made. An example of this was seen in the 
X-ray observations of the type 1.9 Seyfert galaxy {NGC~2992} 
(Weaver et al. 1996) which showed a delay between the Fe K$\alpha$ 
emission line strength and the declining 2 -- 10 keV continuum.

The comparison of the equivalent width of the Fe K$\alpha$ emission
lines to the measured column density in a sample of narrow line AGN
(both type 2 and NELG $\equiv$ type 1.8 and 1.9) observed by $ASCA$
has shown that reprocessing by reflection and scattering can be also
important (Turner et al. 1997ab; 1998). However these measurements did
not indicate a location for the absorbing or reflecting medium because
only a single epoch observation was presented for each object.
Examination of the long term variation in the hard X-ray (2~--~10~keV)
flux from type 2 Seyfert galaxies has shown that changes by a factor
of 2 -- 3 are not unusual (Ueno 1997) and so we may observe a delay in
the response of the Fe K$\alpha$ line to changes in the direct
continuum.

Here we present multiple X-ray observations of two
narrow line AGN, the type 1.9/2 Seyfert galaxy {NGC~4388} and the type
1.9 Seyfert galaxy {ESO~103--G35}, to search for evidence of variations
in the absorbing column, the 0.5 -- 10 keV X-ray flux and Fe K$\alpha$ 
emission line over a period of $\sim$ 2 years.

\subsection{{NGC~4388}}

The first type 2 Seyfert galaxy to be unambiguously detected in hard
X-rays was {NGC~4388}, a nearly edge on spiral galaxy 
(i $\sim 72^{\circ}$, SB(s)b pec), that is the dominant source of hard 
X-rays in the Virgo cluster (Helou et al. 1981; Hanson et al. 1990; 
Takano \& Koyama 1991).  The low spectral resolution of the $SL2$ XRT 
and $Ginga$ LAC observations did not allow
for a firm detection of any emission features in the spectrum of
{NGC~4388}. The $SL2$ spectrum of {NGC~4388} could be modeled as a fairly
flat ($\Gamma \simeq 1.5^{+0.9}_{-0.5}$) power law continuum absorbed
by a large column density of cold gas ($N_{\rm H} = 2.1^{+2.8}_{-1.4}
\times 10^{23}$ cm$^{-2}$; Hanson et al. 1990).  The
H$\alpha$/H$\beta$ emission line ratio of the nuclear emission implies
an extinction of $A_{\rm V} \sim 1$ to 1.5, equivalent to a column
$N_{\rm H} = 5 \times 10^{21}$ cm$^{-2}$ (assuming a standard gas to
dust ratio).  Infrared spectroscopy has also revealed a strong
9.7$\mu$m absorption feature in the spectrum of the nucleus of
{NGC~4388} consistent with the presence of large amounts of obscuring
material (Roche et al. 1991).

High resolution spectra of the nuclear emission shows evidence of
broad H$\alpha$ emission (FWZI~$\sim~6000$~km~s$^{-1}$, Filippenko \&
Sargent 1985) although at a much weaker level than that claimed by
Stauffer (1982). This, along with the detection of broad H$\alpha$
emission from a region a few kpc from the nucleus (Shields \&
Filippenko 1988; 1996), results in a re-classification of {NGC~4388} as
an intermediate type 1.9 Seyfert galaxy.  However, no broad Pa$\beta$
emission has been detected at either the nucleus of {NGC~4388} (Ruiz,
Rieke, \& Schmidt 1994; Goodrich, Veilleux, \& Hill 1994) or the
off-nuclear position where broad H$\alpha$ was detected (Blanco, Ward,
\& Wright 1990). Spectropolarimetry of {NGC~4388} has also failed to 
conclusively reveal broad emission lines in polarized light (Young et
al. 1996) and the continuum polarization is low ($\sim~2$\%), most
likely arising in dusty interstellar medium of the host galaxy (Kay
1994).

The Virgo cluster has been the target of a dozen observations by the
$ASCA$ satellite between 1993 and 1995, {NGC~4388} appeared in the field of
view of three of them. Here we present for the first time the result from a
targeted observation of {NGC~4388} taken in 1995. We also reanalyze
one of the serendipitous 1993 observations previously reported by
Iwasawa et al. (1997, hereafter I97). This is combined with a nearly
simultaneous $ROSAT$ PSPC observation to assist in the determination
of the nature of the soft X-ray emission from {NGC~4388}. We also
search for the presence of hard X-ray reprocessing by studying a
$CGRO$ OSSE observation taken 1 month after the 1993 $ROSAT$ and
$ASCA$ observations.

\subsection{{ESO~103--G35}}

The {$HEAO$~A2} hard X-ray source {1H~1832--653} was identified with the
S0/Sa type galaxy {ESO~103--G35} (z = 0.01329) by Phillips et al.
(1979) who classified the AGN as a type 1.9 Seyfert due to the
weakness of the H$\beta$ emission line and the presence of a broad
H$\alpha$ line with FWZI 6000 km s$^{-1}$. Examination of the ESO (B)
atlas shows that the host galaxy has an inclination angle of $\sim
60^{\circ}$ (Kirkahos \& Steiner 1990). A deep 9.7$\mu$m absorption
feature implies a substantial optical depth at 10$\mu$m toward the
emitting source (Roche et al. 1991) and {ESO~103--G35} is also seen to
host an H$_{2}$O megamaser (Braatz, Wilson, \& Henkel 1996).  A large
absorbing column ($N_{\rm H} > 10^{23}$ cm$^{-2}$) was directly observed in
$EXOSAT$ and $Ginga$ observations (Turner \& Pounds 1989; Warwick et
al. 1993). The $EXOSAT$ observations of {ESO~103--G35} indicated a
possible variation in the column density over a period of four months
interpreted as due to motion of material in the BLR across the 
line-of-sight to the continuum source (Warwick, Pounds, \& Turner 1988). 
We present three $ASCA$ observations of {ESO~103--G35}; note however 
that the 1994 observation was previously reported in Turner et al. (1997a).

\section{Observations}

\subsection{$ASCA$}

The $ASCA$ observations of {NGC~4388} and {ESO~103--G35} examined here
were obtained from the HEASARC archives as well as proprietary data
and analyzed using the XSELECT and XSPEC software packages (Arnaud
1996).  Tables 1a and 1b give a log of the $ASCA$ observations
examined in this paper along with the screening criteria used to
select useful events.  Standard background cleaning was applied to
remove events due to high-energy particles, and the light curves were
examined for data dropout and periods of high background level.
Hot and flickering pixels were removed from the Solid--state Imaging
Spectrometer (SIS) data. The faint mode data were converted to BRIGHT2
mode after correction for dark frame error and echo effect.

The SIS detectors have suffered a steady degradation since launch due
to the harsh radiation environment in orbit. There are two major
effects, Residual Dark Distribution (RDD) and non-uniform Charge
Transfer Inefficiency (CTI), both of which impair the CCD chips in the SIS
instruments (see Dotani et al. 1995; 1996 for details).  These effects
can be corrected for in the extraction of the events from the
detectors, but only the 1995 {NGC~4388} and 1996 {ESO~103--G35} data were
corrected for RDD as the effect is not significant in observations
taken before 1995. Note that only the 2-CCD mode data from the 1993
observation of {NGC~4388} is examined here as the few events which we were 
able to extract from the 4-CCD mode data did not significantly improve the
signal-to-noise ratio of the data.  

{NGC~4388} is in the field of view of both the Gas Imaging Spectrometer
(GIS) and the SIS detectors in only one of the 1993 $ASCA$
observations (of {NGC~4374} on 4 July 1993, hereafter the 1993b
observation) and this observation is examined in detail here.  The
distribution of X-ray events from {NGC~4388} on the detectors during the
remaining serendipitous observation (of {NGC~4406} on 3 -- 4 July 1993,
hereafter the 1993a observation) is somewhat uncertain and the
extreme off-axis position ($18\arcmin$) of {NGC~4388} in the GIS 2 data
and its proximity to the calibration source in the GIS 3 data may give
spurious results. The choice of areas on the GIS and SIS detectors to 
represent an accurate background for the observations of {NGC~4388} is 
problematic as this galaxy lies 1.3$^{\circ}$ from M87 and has two giant 
elliptical galaxies, {NGC~4374} (M84) and NGC~4406 (M86), within 18$\arcmin$. 
This creates a complicated mosaic of X-ray emission across the detectors 
and the extraction regions must be chosen with care to collect as many
events due to the AGN as possible but also to account correctly for
variations in the intensity of the background thermal X-ray emission
that is known to exhibit a variety of temperatures (Takano et
al. 1989; Awaki et al. 1994). For this reason the use of the $ASCA$ 
blank sky observations for background regions is unsuitable even 
though {NGC~4388} lies at high Galactic latitude ($l^{II} = 74^{\circ}$).

Although {NGC~4388} is not a bright X-ray source, extraction regions
6\arcmin \hskip 0.1cm in radius were used, as recommended for GIS
observations of bright point sources. The background extraction region
was chosen to be an annulus around this circle out to a radius of
8\arcmin, different from those employed by I97. However, we find
that the background spectra from the regions chosen by I97 and those
used here are essentially identical.
The extraction regions chosen for the SIS observations of {NGC~4388}
are polygons lying within the chip boundaries, approximating circular
regions of radius 4\arcmin \hskip 0.1cm to 5\arcmin \hskip 0.1cm as
recommended for SIS observations of point sources (Day et
al. 1995). The background was taken from regions around the source
extraction region out to the edge of the same chip but within a radius
of $\sim$ 8\arcmin. The standard technique for background subtraction
of using regions around the edge of the same chip that the target is
placed on was found not to be suitable for the observations of
{NGC~4388} due to the spatially variable nature of the soft X-ray
emission from the Virgo cluster.

The radial profile of the X-ray emission from {NGC~4388} was
investigated in the $ASCA$ observations in both soft X-ray (0.5 - 3.0
keV) and hard X-ray (3.0 - 10.0 keV) bands; however no indication of
extended emission was found. This is not surprising as the emission
observed in the 1991 $ROSAT$ HRI observation that extends out to
$40\arcsec$ (Matt et al. 1994) and the 2\arcmin \hskip 0.2cm X-ray
lobe seen in the 1993 PSPC observation (Colbert et al. 1998) would be
difficult to detect with the $ASCA$ instruments providing $\lesssim 100$
total events above background during the observations.

{ESO~103--G35} was observed by $ASCA$ on three occasions, September
1994, September 1995, and March 1996 (see Table 1b). The standard
extraction region of a circle with a radius 6\arcmin \hskip 0.1cm for the
target events and an annulus out to 8\arcmin \hskip 0.2cm for the
background events were used for the GIS observations of {ESO~103--G35}.
For the SIS observations of {ESO~103--G35} the target events were taken
from circular regions of 4\arcmin \hskip 0.1cm radius centered on
{ESO~103--G35}, except for the 1996 SIS 1 data which were extracted from
a $8\arcmin.4 \times 7\arcmin.2$ box as {ESO~103--G35} was too close to
the edge of the chip for a circular extraction region to be used.
Background events were taken from rectangular regions around the edge
of the same chip with approximately the same area as the extraction
regions, avoiding a regions closer than $\sim$ 5\arcmin.5 to the
target.  The radial profile of all the $ASCA$ observations of
{ESO~103--G35} were consistent with a point source for events above 2
keV, there being too few events below that energy to make a useful
comparison.

The data from the GIS and SIS instruments were rebinned so that each
spectral channel contained at least 20 counts so that chi-square
statistics are applicable for spectral fitting. Standard response
files were used for the GIS data and response matrices were created
for each SIS instrument. Ancillary files for both GIS and SIS data
were also created using FTOOLS packages.  Note that the GIS and SIS
data (both 2 CCD and 4 CCD) are modeled simultaneously including a
parameter needed to account for minor differences in the
normalizations of the GIS and SIS instruments.

\subsection{$ROSAT$ PSPC}

{NGC~4388} has appeared in two $ROSAT$ PSPC observations; however the
first of these (the December 1991 observation of {NGC~4406}) placed
{NGC~4388} in a region on the detector shadowed by the PSPC window
support structure and therefore this observation was not analyzed. 
It was fortunate that a targeted $ROSAT$ PSPC observation of
{NGC~4388} was made during the 1993 $ASCA$ observation. Combining the 
$ROSAT$ and $ASCA$ spectra extends the band pass to lower energies 
where the sensitivity of the $ASCA$ instruments is rapidly declining.  
An analysis of the PSPC spectrum (Rush \& Malkan 1996) and the image 
(Colbert et al. 1998) is available and so we shall only describe the 
results from a simultaneous fitting of the $ROSAT$ and $ASCA$ data (\S 3.8). 
The $ROSAT$ PSPC data were extracted in the standard manner (Turner 1996) 
with periods of high background level excluded. The background subtracted 
count rate between 0.1 and 2.0 keV was 0.045~s$^{-1}$ for a total exposure 
of 11.65~ks.

A $ROSAT$ PSPC observation of {ESO~103--G35} was performed between
31~March and 12~April 1993 (Colbert et al. 1998) with the target
4.$\arcmin$6 off-axis.  A standard extraction aperture of radius
$2\arcmin.5$ was used. A nearby bright source was excluded from the
background region, chosen to be an annulus of outer radius 4\arcmin.  
The final exposure time, after excluding periods of high background, is
16.3~ks giving a 0.1~--~2.0~keV background subtracted count rate of
$4.9~\times~10^{-3}$~s$^{-1}$.  There are two faint point sources,
with PSPC count rates of $\sim~0.03$~s$^{-1}$, within 6\arcmin \hskip
0.1cm of the position of ES0~103--G35 and these will have contaminated
the soft X-ray events observed with the $ASCA$ instruments.  The PSPC
spectra of these two sources can be fitted with a blackbody model of
temperature 0.12 and 0.14 keV.

\subsection{Short time-scale variability}

The light curves of all the $ASCA$ and $ROSAT$ observations of {NGC~4388}
and {ESO~103--G35} were examined for evidence of short term X-ray
variability.  No significant variability was observed in the 
background-subtracted events due to {NGC~4388} during a 1.75 day $ROSAT$ 
HRI observation in December 1991 or during the 1993 $ROSAT$ PSPC 
observation that spanned 16 days.  A comparison between GIS detector 
areas centered on the position of {NGC~4388} and various regions on the 
detector during the 1993 $ASCA$ observations showed no significant 
variability of events due to {NGC~4388} sampled on time-scales from 
minutes to days. The X-ray events were also split into two broad energy 
bands but no variability was discovered in either the soft 
($0.5~<~E~<~3.0$ keV) or hard X-ray events ($3.0~<~E~<~10.0$ keV) during 
the observations. The events from both the GIS and SIS instruments 
during the 1993b and 1995 observations were summed and the background
events subtracted. The energy range of the events were also restricted
to be above 2 keV, where the flux from the AGN dominates over the
Virgo cluster emission, and below the energy where {NGC~4388} is
detected above the background level (10 keV for the GIS and 9 keV for
the SIS observations).  The variations are statistically significant
at the 99\% level only for the 1995 observations; however the uncertainties
in the count rates are large and the GIS and SIS detector variations
do not match very well. Thus we cannot confirm that any variations in
X-ray emission occurred on a time-scale of days during the $ROSAT$ and
$ASCA$ observations of {NGC~4388}.

Statistically significant variations in count rate were observed 
during all the $ASCA$ observations of ESO~103--G35. However we only claim 
significant variability for the 1996 observation where 
$\chi^{2}_{\nu}$ (SIS) = 7.2 and $\chi^{2}_{\nu}$ (GIS) = 5.5 ($\nu$ = 6)
compared to a constant source.
In this observation the signal to noise 
is highest and the variations in count rate are similar in both the GIS 
and SIS detectors and so we present this light-curve in Figure 1.

\section{Spectral Analysis}

\subsection {The $ASCA$ spectra of {NGC~4388}}
 
For each observation a number of different continuum models were
applied to the data, and for each case the chi-square statistic of the
goodness of fit along with the number of degrees of freedom in the
model fit are quoted (the 1993b data followed by 1995). The
F-test probability statistic (Bevington \& Robinson 1992) is also
quoted where different models are compared. An absorbed power law
model (1PL) can be rejected for all the observations at $> 99.99$\%
level ($\chi^{2}/dof$ = 447.2/329; 873.2/541). The residuals from this fit 
to the SIS data are presented in Figure 2.  Immediately obvious in all the
$ASCA$ spectra is a strong emission line near 6.4 keV and a soft X-ray
excess below 3 keV.  The poor statistics seen in the spectra below 3
keV is a consequence of the subtraction of the high background flux
from the Virgo cluster whose thermal spectrum peaks around 1 keV.

When the soft X-ray emission was modeled as an unabsorbed power law
continuum (2PL), it was found that the model fit improved significantly
(393.9/327, $P^{\rm 1993}_{\rm F-Test} \ll 0.001$; 629.0/539, $P^{\rm
1995}_{\rm F-Test} \ll 0.001$).  No improvement was made over the 2PL
model if instead a Raymond-Smith thermal plasma emission model
(Raymond \& Smith 1977) is added to an absorbed power law (RS+PL) to
model the soft X-ray emission.

However the model used by I97 included both thermal emission and
unabsorbed power law components (RS+2PL), and when this model is
applied to the data the chi-square values of the residuals are
significantly reduced in both observations ($\Delta\chi^{2}$/$\Delta
dof$ 9.6/2; 25.0/2). When the photon indices of the power law components 
are then constrained to be the same, the model fit did not significantly
degrade or improve. There was also no improvement when each continuum 
component was associated with a separate column of absorbing material 
and it was found that the columns absorbing the thermal and the (less absorbed)
power law components were consistent with that found in the RS+2PL
model. Models that included 2 thermal components (Raymond-Smith and
Blackbody or 2 Raymond-Smith components) in addition to a heavily
absorbed power law also provided a reasonable fit to the data
(396.7/322; 641.7/534). However the temperature of the second thermal
component was $\sim$ 10 keV and could not be
constrained and so we rejected this model.

We agree then with I97 that the RS+2PL model proves a satisfactory
description of the continuum emission seen in the $ASCA$ observations
of {NGC~4388}.  The addition of a Gaussian emission feature near 6.4 keV
also improves upon the RS+2PL model fit significantly (320.6/321
$P^{\rm 1993}_{\rm F-Test} \ll 0.001$; 542.3/533 $P^{\rm 1995}_{\rm F-Test}
\ll 0.001$) for both observations and will be discussed in \S 3.4
below.  Table 2a presents the RS+2PL models along with the 90\% error
estimate for 4 Interesting Parameters (IP). These models, unfolded
from the instrumental response, are shown in Figure 3 along with the
ratio of the data to the folded model.  The confidence contours for
the parameters in the RS+2PL models 
fitted simultaneously to the SIS and GIS data in each observation
are presented in Figure 4, where
the contour levels are $\Delta\chi^{2}$ at 68\%, 90\%, and 99\% levels
for 4 IP ($\Delta\chi^{2}$ 4.72, 7.78, and 13.3 respectively) except
for Fig. 4f where the contours are $\Delta\chi^{2}$ for 2IP.  This
is a departure from the standard reporting of error estimates for
modeling of X-ray spectra where errors based on 1 or 2 IP are usually
quoted. However, in complicated, multi-component models, each
continuum model parameter is sensitive to changes in other parameters
and so the $\Delta\chi^{2}$ values of more IP should be quoted to give
a more legitimate estimate of the errors in the model (Yaqoob
1998). As an example, the error estimate for the photon index includes
the photon index, the normalizations of both power law components and
the absorbing column as interesting parameters, the other parameters
being frozen at their best fit values.

The total fluxes between 0.5 and 10.0 keV for these models are $f_{\rm
X}$ = (1.42, 0.68) $\times 10^{-11}$ ergs cm$^{-2}$ s$^{-1}$ (1993,
1995 respectively). The power law continuum models shown in Table 2a
give an absorption corrected 2 -- 10 keV luminosity for {NGC~4388} of
$L_{\rm X}$ = (2.78, 1.28) $\times 10^{42}$ ergs s$^{-1}$ (1993, 1995
respectively) for a distance to the Virgo cluster of 27 Mpc (Fabbiano,
Kim, \& Trinchieri 1992).

\subsection{The $ASCA$ spectra of {ESO~103--G35}}

The residuals from fitting an absorbed power law to the $ASCA$
observations of {ESO~103--G35} are presented in Figure 5. Residuals
around 6 keV and below 3 keV can be seen and there is also a feature
near 0.9 keV in the 1994 and 1996 data that may be an emission line,
or blend of lines. The absorbed power law model is a poor description
of the observations; the $\chi^{2}/dof$ for 1994, 1995 and 1996
respectively are 320.2/280, 105.9/98 and 615.6/532. The addition of a
Gaussian emission line near 6.4 keV to the absorbed power law
significantly improves the fit in all the observations (274.7/277,\hskip 0.1cm 
82.8/95 and 575.6/529; {$P_{\rm F-Test}^{\rm 1994} < 0.001, \hskip
0.2cm 0.01 < P_{\rm F-Test}^{\rm 1995} < 0.001$ \hskip 0.1cm and
\hskip 0.1cm $P_{\rm F-Test}^{\rm 1996} < 0.001$}) and is discussed in
detail below (\S 3.4).  A significant improvement is made in the 1994
and 1996 observations when the soft X-ray residuals are accounted for
with the addition of a second (less absorbed) power law (262.7/275,
$0.02 < P_{\rm F-test}^{\rm 1994} < 0.01$ and 534.0/527, $P_{\rm
F-test}^{\rm 1996} < 0.001$) There are very few soft X-ray counts in
the 1995 observation and so the improvement in adding a second power
law is not significant (78.2/93, $0.2 < P_{\rm F-test}^{\rm 1995} <
0.1$) but for consistency we include the second power law in the
analysis of the 1995 observation.

The contaminating sources seen in the $ROSAT$ PSPC observation (\S2.2) should
have provided at least twice as many soft X-ray events than those
observed during the $ASCA$ observations. One of these sources can be
identified with a $m_{\rm B} = 11.5$ star with a proper motion of 
$\mu = 0.165 \pm 0.022$ arcsec yr$^{-1}$ (Wroblewski \& Torres 1994), 
and the other remains unidentified. The flux from the PSPC observation of 
the weak, unresolved nuclear soft X-ray source identified with {ESO~103--G35} 
is consistent with that seen during the 1994 $ASCA$ observation, although
slightly low compared to that observed in the 1996 observation. There
are so few soft X-ray events accumulated during the $ASCA$
observations that a model that combines thermal and power law emission
to describe the flux below 2.5 keV does not allow the parameters to be
meaningfully constrained.  A model where the flux below 3 keV is due
to thermal emission alone does not improve on the two power law model
(261.3/273, 77.7/91 and 535.3/525) and the low signal to noise does
not allow a good constraint of the temperature of this component to be
made (e.g. $kT \hskip 0.1cm {\rm (1994)} \lesssim 1.9$ keV).
We therefore parameterize the soft X-ray emission with a power law
alone noting that a proportion of the flux may be due to contaminating
sources.

The parameters for the two power law plus Gaussian emission line model
are presented in Table 2b, the model unfolded from the
instrumental response and the data to model ratio are shown in
Figure~6, and the $\chi^{2}$ confidence contours of the model 
fitted simultaneously to the GIS and SIS data in each observation are
shown in Figure 7. The total 2~--~10~keV
flux is $f_{\rm X}$ = (1.42, 2.43, 2.41) $\times 10^{-11}$ ergs
cm$^{-2}$ s$^{-1}$ (1994, 1995, 1996 respectively). The
absorption-corrected 2 -- 10 keV luminosity of {ESO~103--G35}
calculated from the power law parameters is found to be $L_{\rm X}$ =
(2.62, 3.11, 4.73) $\times 10^{43}$ ergs s$^{-1}$ (1994, 1995, 1996
respectively) for a redshift of $z = 0.01329$ assuming H$_{0}$ = 50 km
s$^{-1}$ Mpc$^{-1}$ and $q_{0} = 0.5$.

\subsection{Fe K edge}

The possibility of `excess' absorption at 7 keV, i.e. more than that
required for the measured absorbing column, was investigated. A model
with absorption by neutral material of Solar metalicity was applied to
the data but with the Fe abundance set to zero in order to measure the
strength of the absorption edge.  It was found that the presence of an
edge due to {\it neutral} \hskip 0.1cm iron (near 7 keV) is required 
(i.e. significantly improves the model fit) in all the 
$ASCA$ observations of NGC~4388 and ESO~103--G35. The optical depth of the
edge in the observations of NGC~4388 was found to be $\tau_{\rm 93} =
0.42 ^{+0.36}_{-0.26}$ at $E^{\rm edge}_{\rm 93} = 7.26
^{+0.39}_{-0.41}$ keV and $\tau_{\rm 95} = 0.54 ^{+0.35}_{-0.27}$ with
$E^{\rm edge}_{\rm 95} = 7.31 ^{+0.68}_{-0.53}$ keV. For the 1994
observation of ESO~103--G35 we find $E_{\rm edge} = 7.09
< 8.0$ keV and $\tau = 0.16$ ($<$ 0.35). Note that the errors
quoted for the absorption edge parameters are 90\% confidence for {\bf 2 IP}. 
The confidence contours for the absorption edge parameters are presented
in Figure 8 and are consistent with no change in the optical depth
between the observations of NGC~4388.

The exception is the 1996 observation of {ESO~103--G35}.  This
spectrum can be fitted with an edge at $7.37 ^{+0.26}_{-0.22}$ keV
(errors are 90\% confidence for 4 IP) with an optical depth of $\tau =
0.47 ^{+0.16}_{-0.12}$.  This model is a marginally significant improvement
over the neutral absorption edge model ($\Delta\chi^{2}$/$\Delta dof$
= 8.86/2) and is not an artifact of the systematics in the data as the
feature is significant in all 4 detectors. This absorption edge is
consistent with that seen in the 1988 $Ginga$ observation of
{ESO~103--G35} (Warwick et al. 1993). The energy of the edge feature
implies that a significant proportion of the iron is in an ionization
state between \ion{Fe}{9} and \ion{Fe}{17}. Models of the ionization
structure of photoionized gas in this ionization state imply that the
ionization parameter ($\xi$), which can be thought of as the ratio of
ionizing photons to the density of the absorbing material, is in the
range $25 \lesssim \xi \lesssim 320$ (Kallman \& McCray 1982).  When
the 1996 observation is modeled with absorption by ionized material
(XSPEC absori model) it was found that the absorption column rose to
$N_{\rm H} = 3.5 \pm 0.3 \times 10^{23}$ cm$^{2}$, with an ionization
parameter of $\xi = 83 ^{+40}_{-74}$ , and Fe abundance $Z_{\rm Fe} =
0.97 ^{+0.71}_{-0.52} \hskip 0.1cm Z_{\sun}$ \hskip 0.1cm
($\Delta\chi^{2}$/$\Delta dof$ = 6.5/2 compared to the model of absorption by
neutral material).
This column agrees with the measured optical depth of the
ionized Fe K$\alpha$ edge.  Note that the position of the
Fe~K$\alpha$ emission line is near the neutral value which is expected
for this ionization state. The peak energy does not move to a
significantly higher energy until ionization state {$\sim$~Fe$^{19+}$} is
reached. The effect of resonant trapping opacity will 
reduce the number of Fe K$\alpha$ line photons in material with an
ionization parameter in the range $100
\lesssim \xi \lesssim 500$ but is not a significant effect for 
$\xi \lesssim 100$ and so will have little effect here 
(Matt, Fabian, \& Ross 1993). The confidence contours for the
absorption edge parameters are shown in Fig. 8 where it can be seen
that the 1996 edge is stronger than that observed in
the 1994 observation. The 1994 data cannot distinguish between a
neutral or ionized absorption edge whereas a neutral Fe K edge can be
excluded from the 1996 observation at the 99\% confidence level.

It is possible that only a proportion of the gas along the line of sight
to the continuum source is in a non-neutral state. If we apply a dual
absorber model to the 1996 observation of {ESO~103--G35}, with a neutral
and an ionized column, we find that the ionization parameter rises to
$\xi = 290^{+850}_{-260}$ with absorbing columns are $N^{Cold}_{\rm
H}$ = (1.1 $^{+0.8}_{-0.7}$) $\times 10^{23}$ cm$^{-2}$ and
$N^{Ionized}_{\rm H}$ = (1.7 $^{+1.3}_{-1.4}$) $\times 10^{23}$
cm$^{-2}$. The improvement over the model fit given in Table 2b (2PL)
is not significant ($\Delta\chi^{2}$/$dof$ = 0.1/1). The errors quoted
here are 90\% confidence for 2IP, and these parameters are not well
constrained so we do not explore this model any further. 

Material with a high ionization parameter may produce recombination
features around 1 keV (Netzer 1996; Matt, Brandt, \& Fabian 1996) and
there do appear to be residuals near 1 keV 
in the spectrum of {ESO~103--G35} shown
in Fig. 5. Similar features appear in the $ASCA$ spectra of Mkn 3 and
NGC~4507 and are attributed to hydrogen-like oxygen and helium-like
neon respectively (Griffiths et al. 1998; Comastri et al. 1998). 
 In the reflection-dominated X-ray
spectra seen in a few type 2 Seyfert galaxies such as Circinus (Matt
et al. 1996) and NGC~6552 (Reynolds et al. 1994), emission lines in
the $ASCA$ bandpass are expected to be strong and can provide a measure of
the ionization parameter and abundance in the scattering medium
(Netzer, Turner, \& George 1998).
To test the significance of these features a Gaussian emission feature
was added to the models in Table 2b; however in neither the 1994 or
1996 spectra does the addition of a Gaussian emission feature significantly 
improve the model fit (1994 $\Delta\chi^{2}$/$dof$ = 1.4/3; 1996
$\Delta\chi^{2}$/$dof$ = 2.5/3). 

\subsection{Fe K$\alpha$ emission line}

The results of adding a single Gaussian line to the power law
continuum to account for the the strong emission feature near 6.4 keV
in the $ASCA$ spectra of {NGC~4388} and {ESO~103--G35} can be found in
Tables 2a and 2b.  A single, narrow, Gaussian proves to be a
reasonable description of the emission feature in all of the spectra;
note that the rest-frame values for line peak energy are quoted. The
improvement in the $\chi^{2}$ residuals of adding a Gaussian line were
highly significant in all the observations.  The confidence contours
of the emission line parameters are shown in Figs. 4 \& 7 where it can
be seen that the line is marginally resolved (at the 1$\sigma$ level)
in the 1993 {NGC~4388} and 1994 {ESO~103--G35} spectra. In the other
observations the line remains unresolved.

The peak of the emission line in all of the observations is
consistent with an identification with neutral to partially ionized Fe
K$\alpha$ (up to ionization state \ion{Fe}{19}, Turner et al. 1992).
The branching ratios demand that Fe K$\beta$ fluorescence be produced
at a level 0.135 relative to Fe K$\alpha$ (Leahy \& Creighton 1993)
and lines of this strength were searched for near 7.085 keV (rest
frame).  The addition of a line of the correct relative strength and
peak energy resulted in a reduction in $\chi^{2}$ in all but the 1995
spectrum of {ESO~103--G35} where no emission feature at the correct
energy could be found ($\Delta \chi^{2}$/$\Delta dof$ for {NGC~4388} ---
1993~=~1.6/2, 1995~=~0.3/2; {ESO~103--G35} --- 1994~=~2.09/1,
1995~=~0.47/3, 1996~=~4.0/1). However, only in the 1996 observation of
{ESO~103--G35} were these improvements statistically significant.

When the parameters of the second Gaussian emission were left free it
was found that the 1993 SIS spectrum of {NGC~4388} could be fitted with
an additional narrow line centered at 7.02 keV.  However the model fit
was not a significant improvement over the use of a single Gaussian
($\Delta\chi^{2}$/$\Delta dof$~=~1.6/2).  The 1995 observation of
{NGC~4388} could also be fitted with two narrow ($\sigma~=~0.0$~keV)
Gaussians at 6.35~keV and 6.42~keV that mimic the single broader, but
unresolved line, with no improvement in the model fit ($\Delta
\chi^{2}$/$\Delta dof$ = 0.1/3).  The residuals in the 1996 SIS
observations of {ESO~103--G35} also suggest the presence of a second
line at $\sim 7.0$ keV (rest frame), although there is no significant
improvement to the fit when a second line is added ($\Delta
\chi^{2}$/$\Delta dof$ = 0.3/3).

\subsection{Dual absorption models}

A version of the dual absorber model (Weaver et~al.  1994) is the
`leaky-total absorber' favored for the X-ray spectrum of IRAS
04575--7537 (Vignali et~al. 1998). In this type 2 Seyfert galaxy the
measured absorbing column is an order of magnitude too small to
account for the strength of the Fe K$\alpha$ emission line.  An
absorber with $N_{\rm H} \sim 10^{23}$ cm$^{-2}$ covering 36\% of the
continuum source was invoked to explain both the flat continuum and
the strength of the Fe K$\alpha$ emission line in IRAS 04575--7537.
Complimentary evidence for a larger absorbing column in IRAS
04575--7537 is the large optical depth of the Fe K edge ($\tau_{\rm Fe
K} = 0.33^{+0.14}_{-0.13}$). Such a dual absorber model may be
relevant for the X-ray spectrum of {NGC~4388} because the strength of
the Fe K$\alpha$ emission line implies an absorbing column $N_{\rm H}
\sim 10^{24.5}$ cm$^{-2}$ leading to large optical depth for the Fe K
edge ($\tau > 1.0$).  If a similar leaky-absorber model is applied to
the $ASCA$ observations of {NGC~4388} no significant improvement is
made over the RS+2PL model. However the power law becomes closer to
the $\Gamma \simeq 1.9$ value associated with unified models (see \S3.6) 
and the larger absorbing column can help to explain the strength of the Fe
K$\alpha$ line. We present the model parameters in Table 3 and shall
discuss this in \S 4.2 as this model appears to be an reasonable
choice for the X-ray spectrum of NGC 4388.

For the observations of {ESO~103--G35} however, the parameters for a
leaky-absorber model were very poorly constrained even at the
1$\sigma$ level.  The model does not improve on the simpler single
absorbing column model and the covering fraction is found to be low
($<$ 20\%) and so we do not investigate this model further.

\subsection{Compton reflection}

The hard X-ray spectra of many type 1 Seyfert galaxies exhibit a
`Compton reflection hump' in the 20 -- 100 keV band, expected when
X-ray irradiation of cold ($T < 10^{6}$ K) matter with a high covering
fraction occurs (Lightman \& White 1988; George \& Fabian 1991; Nandra
\& Pounds 1994). An investigation of the $Ginga$ spectra of Seyfert
galaxies found that the effects of a reflection component can be
important below 20 keV and this has led to a modification of the
so called ``canonical'' X-ray power law slope from $\Gamma
\equiv 1.7$ (Turner \& Pounds 1989) to $\Gamma \equiv 1.9$ when the
reflection continuum is taken into account (Nandra \& Pounds 1994). More
recently, observations above 50 keV have shown that a Compton
reflection component is present in the average spectrum of Seyfert
galaxies, with very little difference seen between the hard X-ray spectra of
type 1 and type 2 Seyfert galaxies (Zdziarski et~al. 1995).

We searched for the presence of a reflected continuum from both
neutral and/or ionized material in the
observations of {NGC~4388} and {ESO~103--G35}.  In all the observations
 the model fit was not significantly
improved by using a reflection model.  This was also found to be the
case when the second ($\sim$ unabsorbed) power law component was
modeled as a reflection continuum `leaking' through lower column density 
material.  However, these model fits were
also not significantly worse than the power law models, and so cannot
be ruled out on statistical grounds alone. The power law slopes of the
reflection model fits were in general steeper than for the RS+2PL and
2PL models but within the error estimates for these models (Tables
2ab). We do not include the details of the models here because the
parameters of the reflection components are poorly constrained.  The
relative contribution of a reflected continuum was examined in the
observations but no significant changes between the observations could
be determined. The difficulty in determining whether a significant
level of X-ray reflection is present is not only associated with the
problem of measuring the power law slope when the continuum is heavily
absorbed but also to the limit of $ASCA$ detector sensitivity to
$\lesssim$ 10 keV.

Fortunately, an Oriented Scintillation Spectrometer Experiment (OSSE)
observation of {NGC~4388} was made between 25 August and 6 September
1993 and this allows spectral modeling from 0.1 keV to 1000 keV
when combined with the $ROSAT$ and $ASCA$ observations taken 1 
month previously.  OSSE is one of the instruments aboard the 
{\it Compton Gamma Ray Observatory} (see Johnson et~al. 1993 for a
detailed description of the OSSE experiment, performance and data
analysis procedures).  There was also a 1992 OSSE observation of
{NGC~4388} that observed a flux density of ($6.35 \pm 0.58$) $\times
10^{-6}$ photons cm$^{-2}$ s$^{-1}$ keV$^{-1}$ at 70 keV and a
spectral slope of $2.37 \pm 0.22$ (Kurfess 1994; Johnson et~al. 1994).
The steepness of the observed soft gamma-ray slope is typical for
radio-quiet Seyfert galaxies and can be modeled as an exponential
cutoff in the hard X-ray power law continuum. A comparison between the
1992 and 1993 OSSE observations found no significant flux or spectral
variability. The OSSE data is difficult to interpret due to the
proximity of M87 and because of this the background is taken from only
one side of the offset pointings.  The 1993 observations of {NGC~4388}
are presented in Figure 9, along with the model of the simultaneous
fit to the $ROSAT$ PSPC and $ASCA$ observations.  An
exponential cutoff was applied to the power law continuum with the
cutoff energy at 250 keV, chosen to be to match that observed in the
average OSSE spectrum of three radio quiet type 2 Seyfert galaxies
(NGC 4507, NGC 7582, and MCG--5--23--16; Zdziarski et~al. 1995). It
can be seen that this is a reasonable description of the OSSE spectrum
(Fig. 9, solid line).

A reflection model was applied to the $ASCA$--OSSE data with
parameters taken from the average $Ginga$--OSSE spectrum of three type 2
Seyfert galaxies (Zdziarski et~al. 1995). The relative normalization
of the reflection continuum ($\Re$) was fixed at 1.3 times the
normalization of the power law continuum, and the exponential cutoff
for the power law continuum fixed at $E_{c} = 250$ keV. The other
model parameters were taken from Table 2a. It can be seen from Fig. 9
that a reflection model (dotted line) is not a good representation of
the OSSE data. The final model applied to the observations is the
thermal Comptonization model of Sunyaev \& Titarchuk (1980). The
parameters are a plasma temperature of $kT = 30$ and optical depth of
$\tau = 5.1$ taken from the model found to be a good description of
the average OSSE spectrum of 15 Seyfert galaxies (Johnson et~al. 1994)
and this model shown in Fig. 9 as the dashed continuum.  There are
very few counts above $\sim$ 1 MeV and with only 6 data points for the
OSSE observations the parameters for the models cannot be usefully
constrained. However this model appears to have difficulty in
reproducing the observed spectrum above 200 keV.

\subsection{Spectral Variability}

It is immediately obvious on examination of the model parameters in
Tables 2a and 2b, and the corresponding $\chi^{2}$ confidence contours
shown in Figs. 4 \& 7 that there has been a significant change in the
absorbed power law continuum flux in both {NGC~4388} and {ESO~103--G35},
falling by a factor of 2 in the former case and rising by about the
same factor in the latter. This can also be seen in Figure 10 where we
show the effect of applying the 1993 model of the spectrum of {NGC~4388}
to the 1995 data and the 1994 model of the spectrum of {ESO~103--G35} to
the 1996 data. The residuals from these plots clearly show continuum
variability in both objects and emission line variability between the
observations of {NGC~4388}. The confidence contours shown in Fig. 7
indicate a possible change in the equivalent width of the Fe K$\alpha$
emission line between the observations of {ESO~103--G35}.  It can also
be seen that, although the ratio of the normalizations of the unabsorbed
($A_{2}$) to absorbed ($A_{1}$) power law continua in the spectra of
{ESO~103--G35} is $\lesssim 1\%$, it changes between the 1994 and
1996 observations (see Table 2b). This variability is significant at the
90\% level (Fig. 7c).

To investigate the significance of these changes further we applied
the same models as those described in \S 3.1 and \S 3.2 and froze the
model parameters that did not appear to change significantly between
the observations. These were the power law photon index ($\Gamma$),
the emission line width ($\sigma$), both of the absorbing columns
($N_{\rm H}$ (1) and $N_{\rm H}$ (2)) and for {NGC~4388}, the parameters
of the Raymond-Smith component. The fixed parameters were
chosen to be the average value from the observations, weighted by the
relative error of each measurement. The results are
given in Tables 2a,b and the confidence contours for these model fits are
shown in Figure 11 where for clarity we only show the confidence limits
for 4 IP at the 68\% and 99\% levels.

What the confidence contours show is very interesting.  The
probability that the flux level of the absorbed power law component
remained constant between the observations of both {NGC~4388} and
{ESO~103--G35} can be rejected at $\gg 99$\% (Fig. 11b,e).  However we
can see that the equivalent width of the Fe K$\alpha$ emission line
had remained $\sim$ constant between the two observations of {NGC~4388}
(Fig. 11a), implying that the emission line flux had fallen by the same
level as the underlying continuum. This can be seen in Fig. 11b as a
significant (99\% confidence) decrease in Fe K$\alpha$ line flux.
This is not the case for the observations of {ESO~103--G35}, there has
been a marginally significant (90\%) drop in the equivalent width of
the line between 1994 and 1996 (Fig. 11d). This can be accounted for by
the line flux remaining constant while the underlying continuum rose
(Fig. 11e). These effects can also be seen in Fig. 10. Examination of
the confidence contours in Fig. 11c,f shows that there has been no
significant change in the flux level of the unabsorbed power law
continuum (A$_{2}$) in the observations of {NGC~4388} and
{ESO~103--G35}. Note that this is also true of the flux held in the
thermal emission in the spectrum of {NGC~4388} if the parameters for
this component remain free.

\subsection {The simultaneous 1993 $ROSAT$ and $ASCA$ observation of 
{NGC~4388}}

Although the $ROSAT$ PSPC has a low spectral resolution it may be able
to distinguish between thermal and non-thermal models as well as
giving an indication of the presence of absorption edges due to a warm
absorber (Halpern 1984).  The 1993 PSPC spectrum of {NGC~4388} was
investigated by Rush \& Malkan (1996) and it was found that the
absorbing column was significantly higher than the Galactic value and
that an emission feature may also be present at 0.6 keV.  The soft
X-ray flux observed by the PSPC in the 1993 observation is somewhat
less than that observed during the HRI observation in 1991
 but is consistent with the flux measured during the 1993 and 1995 
$ASCA$ observations.

The serendipitous observations of {NGC~4388} in 1993 by $ROSAT$ and
$ASCA$ allowed for a modeling of the X-ray spectrum from 0.1 to 10
keV.  It was found that the model from Table 2a (RS+2PL) could not
account for the emission seen by the PSPC below $E < 0.4$ keV
($\chi^{2}$/$dof$ = 363.9/350).  A second, cooler, thermal component
with $kT = 0.09 \pm 0.09$ keV was added to the model with an absorbing
column found to be at the lower limit set by the measured Galactic
column ($N_{\rm H}^{\rm Gal} = 2.54 \times 10^{20}$; Murphy et al. 1996). This
significantly improved the model fit ($\Delta\chi^{2}$/$\Delta dof =
27.7/4$, $0.01 > P_{\rm F-Test} > 0.001$), the parameters for the
remaining components were found to be consistent with those in Table
2a. This is discussed in \S 4.3.

\section {Discussion}
 
The defining characteristic of the X-ray spectra of type 2 Seyfert
galaxies is the absorption of a power law continuum by columns of cold
gas with $N_{\rm H} \sim 10^{22}$ to $10^{23}$ cm$^{-2}$ (Smith \&
Done 1996; Turner et al. 1997a). Two locations for the absorbing
material are generally considered: a torus of optically thick cold gas
surrounding the active nucleus and the interstellar medium (ISM) of
the host galaxy. The latter may be a reasonable possibility for
{NGC~4388} and {ESO~103--G35} as both AGN lie in highly inclined
spiral galaxies.  The observed narrow line Balmer decrements are
likely to arise in the host galaxy ISM rather than in a unified model
torus as the narrow line region is usually thought of as being more
spatially extended than the torus.  However, the large columns indicated by the
X-ray observations of {NGC~4388} and {ESO~103--G35} are not consistent
with the expected neutral hydrogen column of 2 and 8 $\times 10^{21}$
cm$^{-2}$ (respectively) indicated by the observed optical extinction
($A_{\rm V}$ = 0.97 and 3.72).  The existence of an H$_{2}$O megamaser
with a luminosity of $L_{\rm maser} = 360 L_{\odot}$ in {ESO~103--G35}
(Braatz et~al.  1996) also provides a strong case for the presence of large
amounts of cold molecular gas close to the active nucleus. These
megamasers have been identified as residing in the molecular torus
surrounding the central engine of a number of AGN and can be used to
determine the physical and geometrical properties of the tori
(Taniguchi \& Murayama 1998 and references therein).

$EXOSAT$ observations of {ESO~103--G35} during 1985 appeared to
indicate a decline in the absorbing column density over a period of
$\sim$ 100 days, and this was interpreted in terms of motion of
BLR clouds across the line of sight (Warwick et~al. 1988). 
There is no evidence for variability in the absorbing column from the
$ASCA$ observations of {ESO~103--G35} alone and the column is within
the error of that measured during the 1988 and 1991 $Ginga$
observations ($1.76 ^{+0.41}_{-0.31} \times 10^{23}$ cm$^{-2}$,
Warwick et~al. 1993; $2.2 \pm 0.3 \times 10^{23}$ cm$^{-2}$, Smith \&
Done 1996).  Examination of the $ROSAT$ All-Sky Survey Bright Source
Catalog shows a bright soft X-ray source (1RXS J184538.1--645144; 1.07
$\pm$ 0.09 counts s$^{-1}$) that is 54\arcmin \hskip 0.1cm from the 
position of {ESO~103--G35} and identified with {CPD--64~3950}, a
{M0~$m_{\rm B}~=~10.4$} star. This star may be the
source of a soft X-ray flare event that occurred during the 1988
$Ginga$ LAC observation ($1^{\circ} \times 2^{\circ}$ field of view) 
which is clearly seen in the $Ginga$ spectrum as a soft excess above
the absorbed power law continuum (Warwick et al. 1993, see
Fig. 9). It is unlikely that this source affected the 1995 $EXOSAT$ ME
observations however as the ME detector had a 45\arcmin \hskip 0.1cm FWHM 
field of view.

The absorbing material may not have been in a neutral state
during the 1996 $ASCA$ observation of {ESO~103--G35} and the measured
ionized absorption edge is in agreement with the edge measured from
1988 $Ginga$ observation of $E = 7.68 ^{+0.12}_{-0.32}$ keV (Warwick
et~al. 1993).  An ionized absorption model appears to be a reasonable
description of the 1991 $Ginga$ spectrum of {ESO~103--G35} with a
similar flux level to that observed in the 1996 $ASCA$
observation. This $Ginga$ observation was fitted by a model with $\Gamma =
1.84 ^{+0.08}_{-0.11}$, $N_{\rm H} =$ ($3.55 ^{+0.72}_{-0.56}$) $\times
10^{23}$ cm$^{-2}$, $\xi = 25 ^{+37}_{-16}$, and EW Fe K$\alpha$ = 220
$^{+100}_{-110}$ eV (Smith \& Done 1996), all consistent with the
ionized absorber model of the 1996 $ASCA$ observation presented here
(\S 3.2).  A comparison between the 1988 $Ginga$ spectrum of
{ESO~103--G35} and the 1994 and 1996 $ASCA$ spectra is presented in
Figure 9 along with the model fit to the $Ginga$ observation (Warwick
et~al. 1993). It can be seen that the flux level of the 1996 $ASCA$
observation is also consistent with the 1988 $Ginga$ observation but
that the continuum flux during the 1994 $ASCA$ observation was lower
than that observed in the $Ginga$ observations in 1988 and 1991.

\subsection{Fe K$\alpha$ emission line variability}

A clue to the location of the absorbing material can come from
studying the behavior of the Fe K$\alpha$ emission line compared with
changes in continuum flux. No variation in the equivalent width of this 
line is observed in the spectra of {NGC~4388} and this implies that the line
flux declined with the continuum. Similar behavior was seen between
the $ASCA$ observations of the type 2 Seyfert galaxy NGC~7172
(separated by a year) where the equivalent width of the Fe K$\alpha$
line remained the same while the continuum flux declined by a factor
of 3 -- 4 (Guainazzi et~al. 1998). This implies that the line emitting
region is within a few light years of the continuum source in these
objects. The $Ginga$ observations of {ESO~103--G35} have indicated an
Fe K$\alpha$ emission line with an equivalent width of $\sim 250$ eV
(Warwick et~al. 1993; Smith \& Done 1996) similar to that observed in
the 1996 $ASCA$ observation. However the equivalent width of the
emission line during the 1994 $ASCA$ observation appears larger than
during these observations, and this becomes a significant difference
when the fixed parameter model is used (Table 2a, Fig. 11d) indicating
 that the line flux remained $\sim$ constant between the two
$ASCA$ observations (Fig. 11e). This behavior may be similar to that
observed in the intermediate type Seyfert galaxy {NGC~2992} and the
type 2 Seyfert galaxy Mkn 3 (Weaver et~al. 1996; Griffiths et~al. 1998),
although in an opposite sense with the line flux increasing at a
slower rate than the continuum for ESO~103--G35.  
The measured column densities in the line of sight material towards
NGC 4388 and ESO 103--G35 could yield an emission line with an
equivalent width of $\sim$ 150 eV for a model where absorption is from
a uniform shell of material entirely covering the continuum source
(Leahy \& Creighton 1993; Awaki et~al. 1991).  The equivalent width of
the Fe~K$\alpha$ emission line for the ionized absorption model of the
1996 observation of {ESO~103--G35} is 225 eV and this can be accounted
for by fluorescence in the line of sight material to within the accuracy of the
measurement of the column density. However the large equivalent widths
measured in the 1994 observation of {ESO~103--G35} and the
observations of {NGC~4388} imply a source of line photons other than
from the line of sight absorbing material and there are a number of
possibilities available to explain this disagreement.

\noindent $\bullet$ The assumptions of the uniform shell model are incorrect. 
For instance the column density along our line of sight may be lower than the
average, or the X-ray emission may be beamed (e.g. Awaki
et~al. 1991). The beaming factor would only have to be a modest factor
of $\sim$ 3 to give the correct line strengths but this would not
account for the change in the equivalent width of the line seen
between the $ASCA$ observations of {ESO~103--G35} unless the
beaming parameters had also varied.

\noindent $\bullet$ If the absorbing material had the toroidal configuration
favored by many unified models of AGN then the excess line photons may
be produced from a scattered or reflected component, perhaps from the
inner edge of the torus or material out of the plane of viewing (case
b in Awaki et~al. 1991). This type of model favors a viewing angle
close to the opening angle of the torus for {ESO~103--G35} as the
expected strength of the scattered flux observed in the soft X-ray
region is $\sim$ 1/1000 that of the primary flux, and this is of the same 
order of magnitude as the measured value of $A_{2}$/$A_{1}$ (Table 2b). 
The higher relative strength of the `unabsorbed' continuum in {NGC~4388} 
(Table 2a) would thus imply a more edge-on viewing angle of the torus 
as a greater proportion of the primary radiation would be obscured. Such a
reflected or scattered continuum may not have had time to respond to
the observed changes in the power law continuum flux, and this is
supported by the lack of a significant variation in the observed
fluxes of this component in {NGC~4388} or {ESO~103--G35}
(Fig. 11c,f). These observations then imply that the inner edge of the
torus would lie $\sim$ 1.5 or 2 light years from the continuum source
in {ESO~103--G35} and {NGC~4388} respectively.

\noindent $\bullet$ An abundance of iron at a factor of $>$ 10 times 
Solar in the absorbing material observed in the $ASCA$ spectra of
{ESO~103--G35} and {NGC~4388} would be required to account for the
high equivalent widths (George \& Fabian 1991) and this would also
cause the Fe K edge to be much stronger than is observed (see \S 3.3).

\noindent $\bullet$ The high column density material in the
leaky-absorber model for the type 2 Seyfert galaxy IRAS 04575--7537
was interpreted as arising in BLR clouds (Vignali
et~al. 1998). However in that case the covering fraction was
significantly lower (36\%) than seen in {NGC~4388} (\S 3.5 and Table 3) 
and much closer to the 10\% value normally associated with the BLR
in AGN. What we may be seeing in {NGC~4388} is the presence of
clumping in a unified model type torus, although we can not rule out the
ISM of the host galaxy of {NGC~4388} as the location of the absorbing material
because the stellar disk of {NGC~4388} has a high inclination to the line of 
sight. The optical depth required to account for the strength of the 
Fe K$\alpha$ emission line in NGC 4388 ($\tau_{\rm Fe K} \sim 1$) is 
also at the upper limit of the 99\% error margin on the measured value 
as can be seen in Fig. 8.
Similarly we can interpret the presence of a less absorbed power law
continuum in NGC 4388 as evidence that the covering fraction of the
lower column density absorber is less than unity (90 -- 95\%),
although this is difficult to reconcile with a model of absorption in
the ISM.  The leaky-absorber model applied to the $ASCA$ spectra of
{ESO~103--G35} causes the power law continuum to become rather steep
($\Gamma \gtrsim 2.2$) and, although the presence of a larger
absorbing column will reconcile the high equivalent width of the Fe
K$\alpha$ emission line seen in the 1994 spectrum, it predicts an
equivalent width higher than is seen in the 1996 spectrum. This model
also does not seem to agree with the variability behavior of the Fe
K$\alpha$ emission line and Fe K edge. The optical depth of the Fe K
edge is seen to increase between the observations and this implies
that the strength of the Fe K$\alpha$ emission line should have
increased but the opposite behavior is seen (Figs. 8 and 11).

\subsection{No reflection continuum in NGC~4388 or ESO~103--G35}

In many type 1 Seyfert galaxies strong Fe~K$\alpha$ emission lines are
observed, some with complex profiles and large equivalent widths
(Nandra et~al. 1997). The interpretation is that of fluorescence from
an accretion disk, however the material in this
case is close to (or coincident with) the source of the continuum
photons and so no lag in line response is expected (e.g. MCG
--5--23-16, Weaver, Krolik, \& Pier 1998; Nandra et~al. 1997).  Turner
et~al. (1998) found evidence for broad Fe K$\alpha$ line emission in
up to a third of their sample of narrow-line AGN which they modeled
as emission from an accretion disk viewed `pole-on' (i.e. similar to
emission seen in type 1 AGN). No evidence for a significant broad line
is found here for {NGC~4388} or {ESO~103--G35}. The identification of
a broad Fe~K$\alpha$ line in the 1994 $ASCA$ spectrum of
{ESO~103--G35} by Turner et~al. (1998) at the 90\% confidence level,
with $\sigma = 310$ eV and $EW = 505$ eV, differs from the results
presented here of a single {\it unresolved} narrow emission line (see
Table 2b and Fig. 7). This could be due to differences in the continuum
model, Turner et~al. (1998) finding a flatter power law and a lower
column density although the measurements presented here agree to
within their 90\% confidence limits. Four of the sample of narrow line
Seyfert galaxies studied by Turner et~al. (1998) have also been
successfully modeled by Weaver \& Reynolds (1998) with a combination
of a broad component from an accretion disk viewed at an inclination
angle of $\sim 48^{\circ}$ and a narrow unresolved component. This
illustrates the difficulty of identifying broad emission features on
heavily absorbed power law spectra with low
signal-to-noise data. 

In type 2 Seyfert galaxies where the power law continuum slope is very
flat and the observed equivalent widths of the Fe K$\alpha$ line are
extremely large, a reflection continuum is thought to be the dominant
component in the hard X-ray continuum. This is the case for
{NGC~1068}, where $ASCA$ observations have revealed a complex blend of
neutral and ionized Fe K$\alpha$ emission lines with equivalent widths
$\gtrsim 1$ keV (Ueno et~al. 1994) and in the Circinus galaxy,
{NGC~6552}, and {NGC~7674}, where equivalent widths larger than 1 keV
are also seen (Matt et~al. 1996; Reynolds et~al. 1994; Malaguti
et~al. 1998 respectively). Nevertheless, the presence of a reflection component
will increase the observed strength of the Fe K$\alpha$ emission lines
as well as cause a flattening of the power law continuum,
although a flat hard X-ray spectral slope
does not necessarily require a dominant reflection component, as is the
case for {NGC~5252} (Cappi et~al. 1996). In this type 2 Seyfert galaxy
the equivalent width of the Fe K$\alpha$ emission line is very low
($\sim$ 100 eV), and although the combination of a partially covered
power law continuum and a reflection component can account for the
flat spectrum ($\Gamma = 1.45 \pm 0.23$), the model predicts an
emission line of equivalent width $\sim$ 400 eV, much larger than is
observed. This situation may also be relevant for {NGC~7582} where a very
flat hard X-ray power law continuum is identified ($\Gamma =
0.7^{+0.3}_{-0.4}$) absorbed by a column $N_{\rm H} = 5^{+2}_{-1}
\times 10^{22}$ cm$^{-2}$ (Schachter et~al. 1998). However only an
upper limit of a few 100 eV could be placed on the equivalent width of
an Fe K$\alpha$ line and so no reflection continuum is likely to be
present. The intermediate strength Fe K$\alpha$ emission line observed
in {NGC~4388} may then indicate that a reflection component is present
but does not dominate the hard X-ray continuum.

The combined $ASCA$ and OSSE spectrum of NGC~4388
is also similar to that of NGC 4507 and NGC 7172 where OSSE
observations combined with $Ginga$ and $ASCA$ observations
(respectively) of these AGN do not require a reflection component
and can be fitted with exponentially cut off power law continua
(Bassani et~al. 1995; Ryde et~al. 1997).  Although both of these type
2 Seyfert galaxies have a heavily absorbed flat power law continuum,
the Fe K$\alpha$ emission line equivalent widths of $\sim$ 300 and 100
eV respectively are consistent with fluorescence through the line of
sight absorbing columns (Comastri et~al. 1998; Guainazzi et~al. 1998)
and so no reflection component is required. The $ASCA$ observations of
NGC~7172 separated by one year also showed that while the continuum
flux declined by a factor of 3--4 the equivalent width of the Fe
K$\alpha$ line remained the same (Guainazzi et~al. 1998). This is
similar to the behavior seen in the $ASCA$ observations of {NGC~4388}
presented here.

For the $Ginga$ observations of {ESO~103--G35}, Smith \& Done (1996)
find that a model with a reflection component is as good a fit as the
ionized absorption model. The reflection model required that $\Gamma =
2.19 ^{+0.09}_{-0.15}$, $N_{\rm H} = $($2.42 ^{+0.26}_{-0.23}$) $\times
10^{23}$ cm$^{-2}$, $\Re = 2.00 ^{+0.48}_{-0.95}$, with an Fe
K$\alpha$ equivalent width of $< 175$ eV. When a reflection model was
applied to the 1996 $ASCA$ spectrum a similar result was obtained
with no improvement over the model presented in Table 2b
($\chi^{2}$/$dof$ = 532.6/524) however, the reflection model
parameters could not be constrained due to the limit of the bandpass
of the $ASCA$ instruments. 
The lower equivalent width of
the Fe~K$\alpha$ emission line seen during 1996 $ASCA$ observation of
{ESO~103--G35} does not require a reflection component and can be
accounted for by line of sight fluorescence in material with the
measured column density alone.

We conclude that the $ASCA$ observations of both NGC~4388 and ESO~103--G35 
do not require a reflection component.

\noindent {\bf NOTE ADDED IN PROOF.} A stronger constraint is found from the
October 1996 and 1997 BeppoSAX observations of {ESO~103--G35} where a
reflection component is indicated in a spectrum that extends up to
$\sim$ 200 keV (Wilkes et~al. 1999, in preparation).

\subsection{Soft X-ray emission}

Significant low energy emission lines identified with elements lighter 
than iron are also present in the spectra of reflection dominated
Seyfert galaxies as predicted in models of reflection from cold
material (Reynolds et~al. 1994; Netzer et~al.  1998). If a reflection
component were strong in the X-ray spectra of either {NGC~4388} or
{ESO~103--G35} then low energy emission features may also be
present. Determination of such features in the spectrum of {NGC~4388} is
complicated by the extended thermal emission seen in the $ROSAT$ HRI
observation (Matt et~al. 1994).  The model of the combined $ROSAT$
PSPC and $ASCA$ spectrum of {NGC~4388} gives a soft X-ray (0.2 -- 4.0
keV) flux of $f_{\rm X} = 7.8 \times 10^{-13}$ ergs cm$^{-2}$ s$^{-1}$
and this is consistent with the total fluxes seen in the 1991 HRI
observation of (6.75 $\pm$ 0.72) $\times 10^{-13}$ ergs cm$^{-2}$ s$^{-1}$
and in the 1978 $Einstein$ IPC observation of $6.6 \times 10^{-13}$
ergs cm$^{-2}$ s$^{-1}$, (Matt et~al. 1994; Fabbiano, et~al.
1992). The radial profile of the soft X-ray emission,
observed by the HRI to extend out to $\sim$ 4.5 kpc, suggests that
$\sim$ 20\% of this flux is from a nuclear component within a radius
1.5 kpc.  The expected 2 -- 4 keV luminosity from the
host spiral galaxy of the AGN in {NGC~4388} is $L_{\rm X} = 6.17 \times
10^{39}$ erg s$^{-1}$, scaling from the B-band luminosity 
(Fabbiano et~al. 1992).  Combined with the observed kiloparsec-scale emission
(Colbert et~al. 1998) this can fully account for the thermal emission
observed in the $ASCA$ spectra. Netzer et~al. (1998) have modeled
emission features below 3 keV in the 1993 $ASCA$ spectrum of {NGC~4388}
as due to a scattering process, however these features may be due to
the thermal component. The addition of Gaussian emission lines to
the spectrum between 1.5 and 3.0 keV does not improve the chi-square
residuals significantly and a scattered continuum cannot explain the
nature of the bulk of the soft X-ray emission seen in {NGC~4388}. The
temperature of the thermal components measured in the combined $ROSAT
- ASCA$ observation are too cool to be identified with a typical soft
X-ray spectrum from a starburst disk (Ptak 1997) or ram pressure
stripped hot gas (e.g. Irwin \& Sarazin 1996; Hwang \& Sarazin
1996). The combined emission from a collection of LMXB's in the bulge
of {NGC~4388} may prove to be the most straightforward explanation of the
nature of the extended soft X-ray emission (Irwin \& Sarazin 1998).
The high spatial resolution of a $Chandra$ observation should prove useful 
in distinguishing between
the various possibilities.

The $ROSAT$ PSPC image of {ESO~103--G35} shows only a weak
unresolved nuclear soft X-ray source with a 0.5 -- 2.0 keV flux of $5
\times 10^{-14}$ ergs cm$^{-2}$ s$^{-1}$. This is consistent with the
flux observed during the 1994 $ASCA$ observation and this poses a
problem. Two nearby soft X-ray sources (see \S 2.2) should have
contaminated the GIS observations with a combined soft X-ray flux at
least 4 times larger than that observed from {ESO~103--G35} during the
PSPC observation. Although the soft X-ray flux appears to have risen
slightly ($9 \times 10^{-14}$ ergs cm$^{-2}$ s$^{-1}$) in the 1996
$ASCA$ observation this is still too low and this may either indicate a
problem with the cross calibration between the $ROSAT$ and $ASCA$
instruments or variability of the contaminating sources.  
The increase in soft X-ray flux
between the 1994 and 1996 observations may indicate an emergence of
scattered or leaked nuclear flux that adds to the soft X-ray emission
from the host galaxy and this could account for the rise in
$A_{2}$/$A_{1}$ when the absorbed power law continuum also rose by a
factor of 2 between the observations. There is also an emission
feature near 0.9 keV in the spectrum of {ESO~103--G35} (Fig. 5) that
may be similar to that seen in the soft X-ray spectra of NGC~4507 and
Mkn 3 (Comastri et~al. 1998; Griffiths et~al. 1998). This was
interpreted as \ion{Ne}{9} K$\alpha$ recombination from a photoionized
plasma in the case of NGC~4507 and this may also be the case for
{ESO~103--G35}. Again, the capabilities of $Chandra$ should resolve these
possibilities.

\section{Summary and conclusions}

$ASCA$ observations of two intermediate type Seyfert galaxies NGC~4388
and ESO~103--G35 have been investigated and combined with $ROSAT$ PSPC
and $CGRO$ OSSE observations. 

The $ASCA$ spectrum of NGC 4388 can be described by a heavily absorbed
flat power law continuum ($\Gamma = 1.4, N_{\rm H} = 3.3 \times
10^{23}$ cm$^{-2}$) with a strong Fe K$\alpha$ emission line ($EW =
720$ eV) and soft X-ray thermal emission ($kT = 0.17$ keV). The data
also require the presence of a second, less absorbed, power law
continuum at $\sim$ 5\% the flux level of the absorbed continuum. 
The equivalent width of the
Fe K$\alpha$ emission line requires a larger column of absorbing
material than is measured and the power law continuum is flatter than
the canonical Seyfert galaxy spectral slope. Both of these properties
are characteristic of reflection dominated type 2 Seyfert
galaxies. However the model fit is not improved by the inclusion of a
Compton reflection component, and there is also a concern that the 
modeling of the contemporary 1993 $ASCA$ and OSSE observations do not 
appear to require a reflection continuum and can be fitted with a flat 
power law with an exponential cut off at high energies. NGC 4388 may
be a case where there is a relatively small amount of reflection that
manifests itself in a stronger than expected Fe K$\alpha$ emission
line. NGC 4388 would then lie somewhere between NGC 5252 and the
Compton-thick type 2 Seyfert galaxies in terms of the relative
strength of the reflection component to the direct absorbed continuum.

A leaky-absorber model is able to reconcile the strength of the Fe
K$\alpha$ line and the `low' measured absorbing column. In this model
there are two distinct absorbing media, the first has a column of
$N_{\rm H} = 5 - 10 \times 10^{23}$ cm$^{-2}$ and a covering fraction
of $\sim$ 70\% and the second a lower column of $N_{\rm H} = 3 \times
10^{23}$ cm$^{-2}$. The strength of the Fe K$\alpha$ line is then
fully accounted for; however the observed optical depth at the Fe K
edge is somewhat smaller than expected from material with such a large
column and the model is unable to improve statistically on the
absorbed flat power law continuum model.  The high column density gas
component in the leaky-absorber model may reside in a hidden BLR of
NGC 4388 (although the covering fraction is higher than normally
associated with BLR clouds) a molecular torus surrounding the active
nucleus, or the host galaxy ISM. The $ASCA$ observations are not
sufficient to distinguish between these possibilities.
The nature of the soft X-ray thermal emission seen in the $ASCA$
observations of NGC 4388 is consistent with an identification with the
extended emission seen in a $ROSAT$ HRI observation. However, higher
spatial and spectral resolution will be required to investigate the
true nature of the extended emission.

The spectrum of the intermediate type Seyfert galaxy ESO~103--G35 can
be well characterized by a heavily absorbed power law continuum of
slope $\Gamma = 1.8$ and $N_{\rm H} = 2.0 \times 10^{23}$
cm$^{-2}$. This continuum was observed to increase in flux by a factor
of 2 between the $ASCA$ observations separated by 18 months. However
the equivalent width of the Fe K$\alpha$ emission line was seen to
decrease between these observations implying that the line flux
remained $\sim$ constant. The optical depth of the Fe K edge is also
observed to increase between the $ASCA$ observations and the edge
energy in the later observation indicates absorption from
partially ionized iron. This matches the results of the 1998 and 1991
$Ginga$ observations (Warwick et al. 1993; Smith \& Done 1996). These,
and the $EXOSAT$ observations taken between 1983 and 1985 (Turner \&
Pounds 1989), also showed ESO 103--G35 to have a similar 2 -- 10 keV
luminosity to that seen in the 1996 $ASCA$ observation.

A Compton reflection component can be included in the continuum model
but is not required on statistical grounds or needed to steepen the
power law continuum to comply with the canonical Seyfert galaxy
spectral slope. It is also not required to reconcile the Fe K$\alpha$
emission line strength in the 1996 observation as this is consistent
with fluorescence from partially ionized material with the measured
column density. However the strength of the line in the 1994 $ASCA$
observation indicates that an additional source of line photons is needed
but a dual absorber model is inconsistent with the observed
optical depth of the Fe K edge. One (highly speculative) possibility
is that the Fe K$\alpha$ emission line is responding to a higher X-ray
flux level previous to the 1994 $ASCA$ observation.

In terms of a unified model of Seyfert galaxies these intermediate
type AGN exhibit an X-ray spectrum very similar to that seen in many
type 2 Seyfert galaxies but are certainly not reflection dominated and
do not require the addition of a reflection continuum to explain the
properties observed in the $ROSAT$, $ASCA$ and OSSE observations
presented here.

\acknowledgements

This paper is contribution No. 677 of the Columbia Astrophysics
Laboratory. KML gratefully acknowledges support through NAG5-3307
(ASCA). This research has made use of the NASA/IPAC Extragalactic 
Database (NED) which is operated by the Jet Propulsion Laboratory, 
California Institute of Technology, under contract with the National 
Aeronautics and Space Administration. This research has also made use 
of data obtained from the High Energy Astrophysics Science Archive 
Research Center (HEASARC), provided  by NASA's Goddard Space Flight Center.

\vfill
\eject

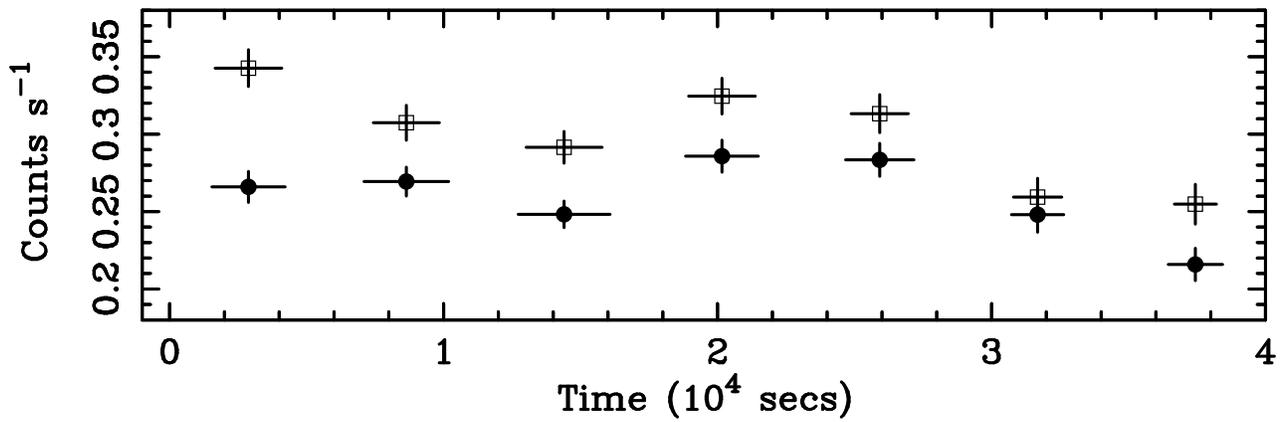
\noindent Fig. 1 --- The 2 -- 10 keV light curve for the 1996 $ASCA$ 
observation of ESO~103--G35. The data is background subtracted and
summed over a full orbit (5760s). $\bullet$ --- Summed GIS detectors,
$\Box$ --- Summed SIS detectors. The length of the horizontal error
bars is a measure of the fraction of GTI within a 5760 second bin.

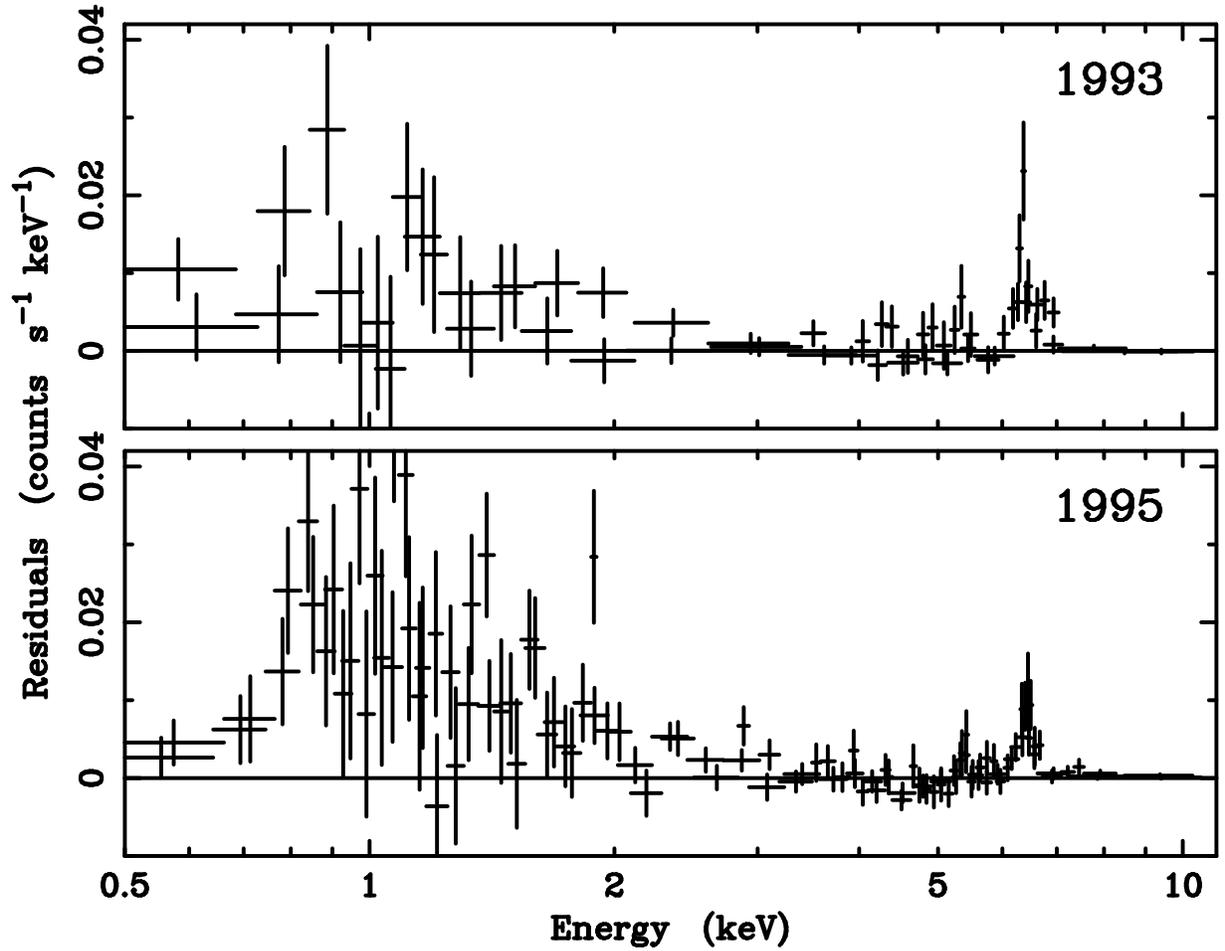
\noindent Fig. 2 --- Residuals from an absorbed power law continuum
model fit to the 1993 and 1995 $ASCA$ spectra of NGC 4388. Only the
SIS data are shown for clarity. Residuals near 6.4 keV suggest the
presence of an emission line and there is a significant `soft
excess'.

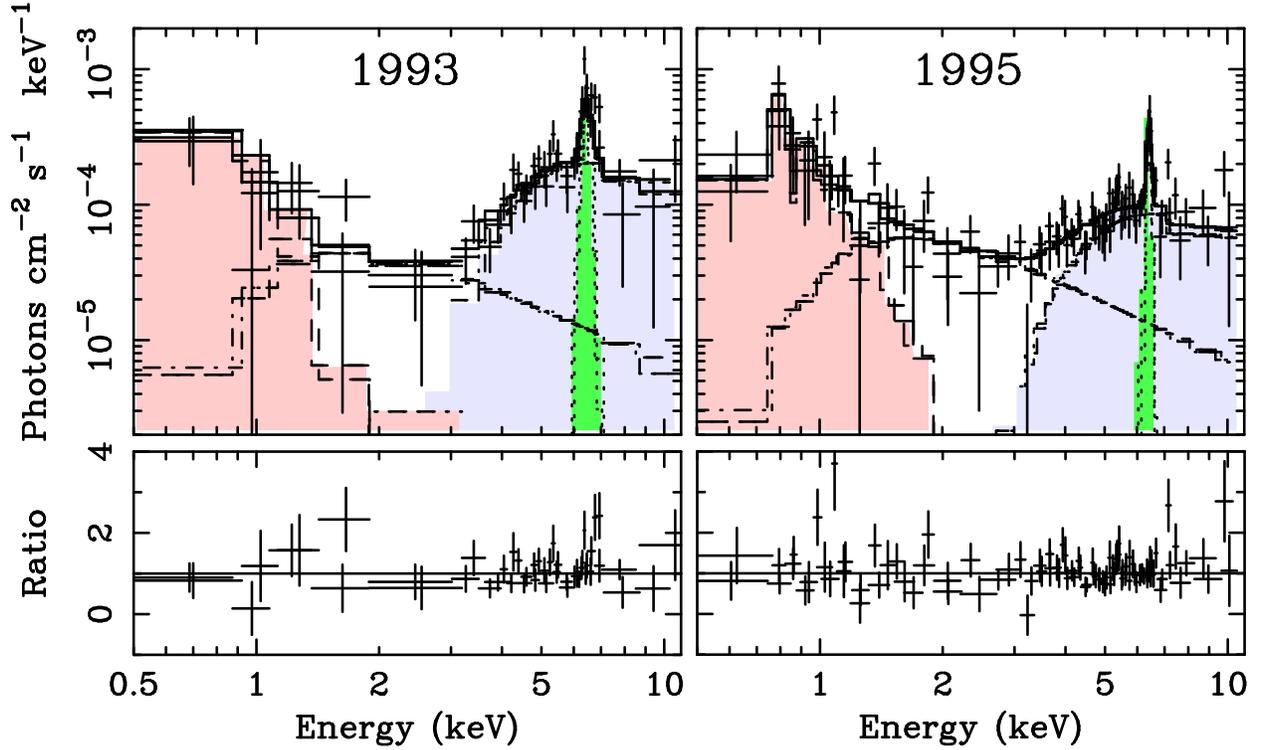
\noindent Fig. 3 --- $ASCA$ spectra of NGC 4388 in 1993 and 1995.
Only the SIS data are shown and have been rebinned for clarity.
{\it Top panels} --- The data and model unfolded from the instrumental
response. The shaded areas represent the Raymond--Smith thermal plasma
emission ($E~\lesssim~2$~keV), the absorbed power law continuum emission
($E~>~3$~keV), and the darker shaded region marks the
Fe~K$\alpha$ emission line. The unshaded component is the less
absorbed power law continuum. {\it Bottom panels} --- The ratio of the
model to the data. 

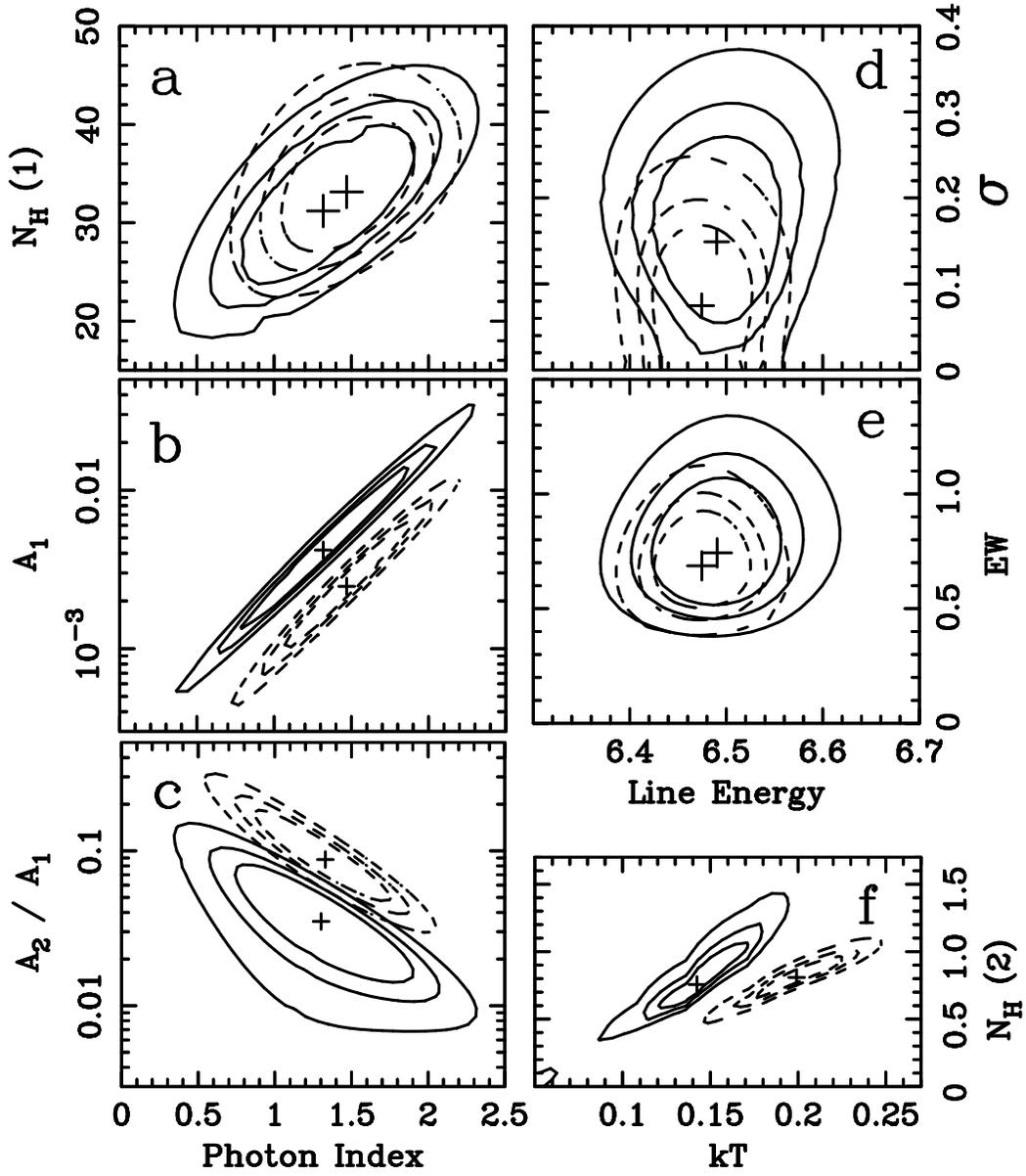
\noindent Fig. 4 --- Confidence contours for the model parameters of the
simultaneous fit to the GIS and SIS $ASCA$ spectra of NGC~4388 in 1993
and 1995 (see Table 2a). Contours represent $\Delta\chi^{2}$
confidence limits (68\%, 90\%, and 99\%) for 4 interesting parameters
(IP). {\it Solid line} -- 1993 observation, {\it dashed line} -- 1995
observation. (a)~Power law photon index vs. absorbing column
($10^{22}$ cm$^{-2}$). (b)~Power law photon index vs. absorbed power
law continuum normalization (photons cm$^{-2}$ s$^{-1}$
keV$^{-1}$). (c)~Power law photon index vs. ratio of the
normalizations of the unabsorbed to absorbed power law continuum
(A$_{2}$/A$_{1}$). (d)~Emission line peak energy (keV) vs. emission
line width (keV). (e)~Emission line peak energy (keV) vs. equivalent
width of the line (keV). (f)~Temperature of the Raymond-Smith
component (keV) vs. second absorbing column ($10^{22}$
cm$^{-2}$), here the contours are $\Delta\chi^{2}$ for 2 IP.

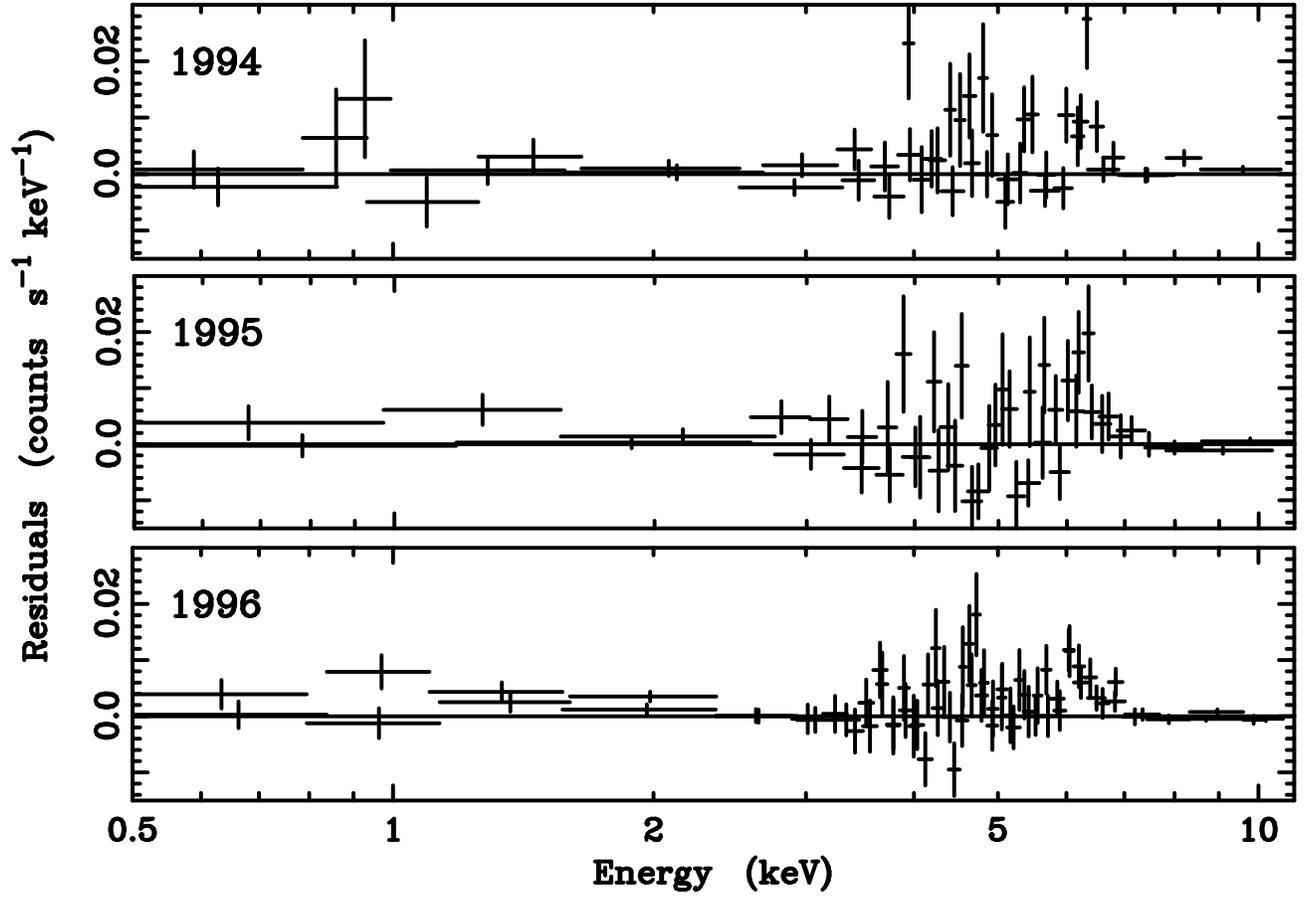
\noindent Fig. 5 --- Residuals from an absorbed power law continuum
model fit to the 1994, 1995, and 1996 $ASCA$ spectra of ESO~103--G35.
Only the SIS data are shown for clarity.

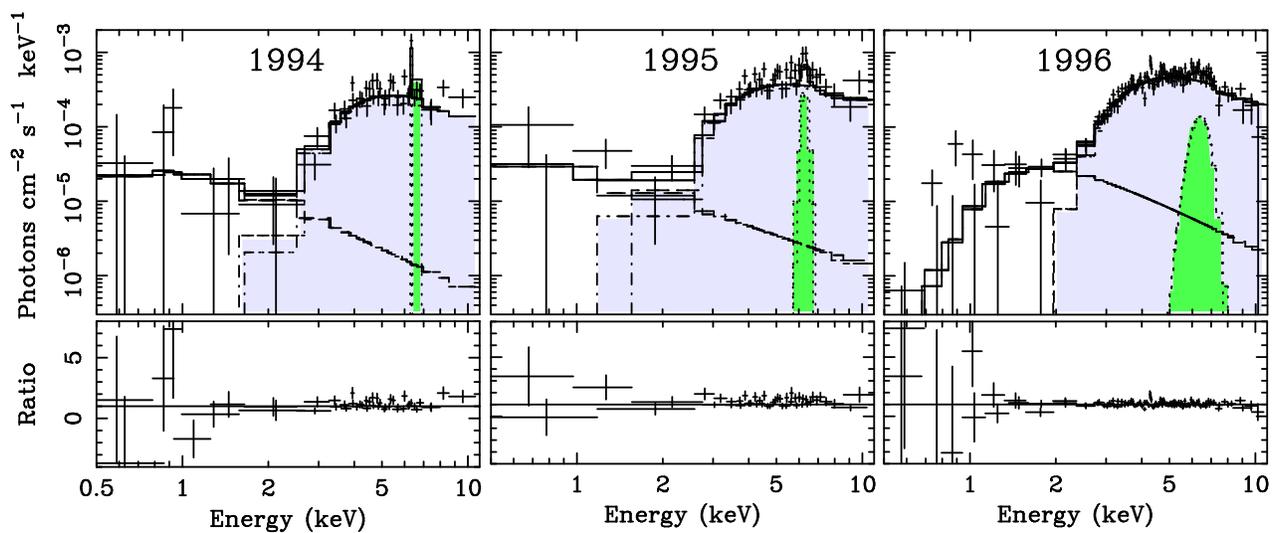
\noindent Fig. 6 --- $ASCA$ spectra of ESO~103--G35 in 1994, 1995, and
1996. Only the SIS data are shown and have been rebinned
for clarity. {\it Top panels} --- The data and model unfolded from the
instrumental response. The shaded area represents the absorbed power
law continuum emission, and the darker shaded region marks the
Fe~K$\alpha$ emission line. {\it Bottom panels} --- The ratio of the
model to the data. 

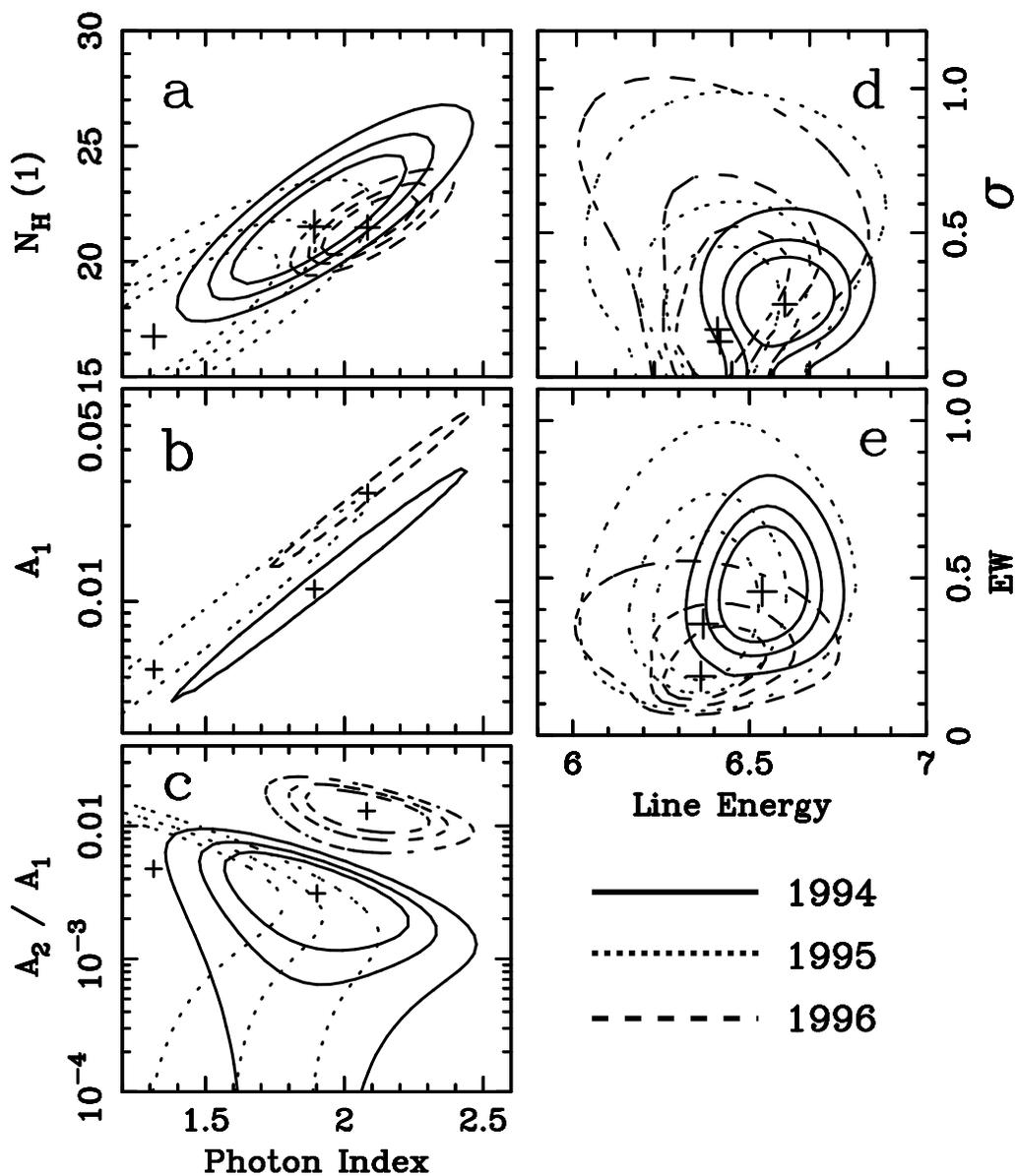
\noindent Fig. 7 --- Confidence contours for the model parameters of the
simultaneous fit to the 1994, 1995 and 1996 $ASCA$ GIS and SIS spectra
of ESO~103--G35 (see Table 2b). {\it Solid line} -- 1994 observation,
{\it dotted line} -- 1995 observation, {\it dashed line} -- 1996
observation. (a~--~e)~See caption for Fig. 4. Note that only the 99\%
confidence contours are shown in panel b for clarity.

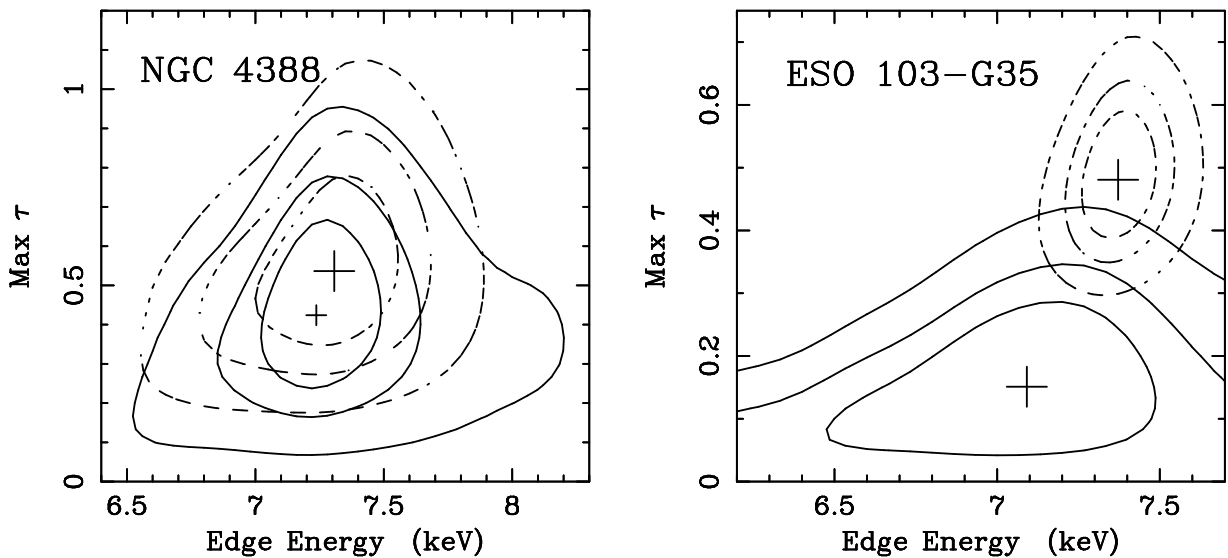
\noindent Fig. 8 --- Confidence contours for the model parameters of the Fe K
absorption edge in the $ASCA$ spectra of NGC~4388 and
ESO~103--G35. Contours are 68\%, 90\% and 99\% confidence limits for 2
IP. \quad {\it Solid lines} -- 1993/94 observations (NGC 4388/ESO
103--G35).  \quad {\it Dashed lines} -- 1995/96 observations. The best
fitted value for the 1993 observation of NGC 4388 is represented by a
smaller cross.

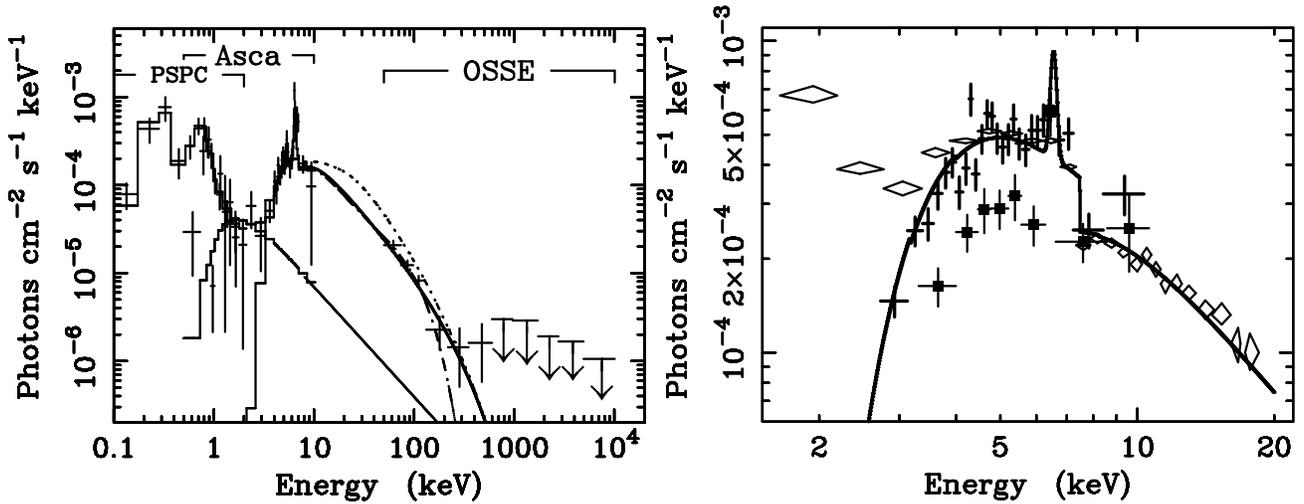
\noindent Fig. 9 --- $Left$ --- The X-ray spectrum of NGC 4388 from 0.1 to
10$^{4}$ keV. The $ROSAT$ PSPC and $ASCA$ SIS model (\S3.8) is used
with an exponential cutoff (solid line) to extend out to the 1993
$CGRO$ OSSE observation. The dotted line is a Compton reflection
spectrum and the dashed line (lower line) is a thermal Comptonization
model. All $ASCA$ instruments
were modeled simultaneously with the PSPC data, but only the rebinned data
from the SIS 0 instrument is shown for clarity.  $Right$ --- The 1988
$Ginga$ observation of ESO~103--G35 (diamonds) shown with the 1994
(black square) and 1996 (crosses) $ASCA$ SIS 0 observations. The model
is that for the $Ginga$ observation with an ionized Fe K edge (Warwick
et al. 1993). The rise in the $Ginga$ spectrum below 3 keV has been
attributed to a contaminating soft X-ray source in the field of view
of the $Ginga$ LAC (Warwick et al. 1993).

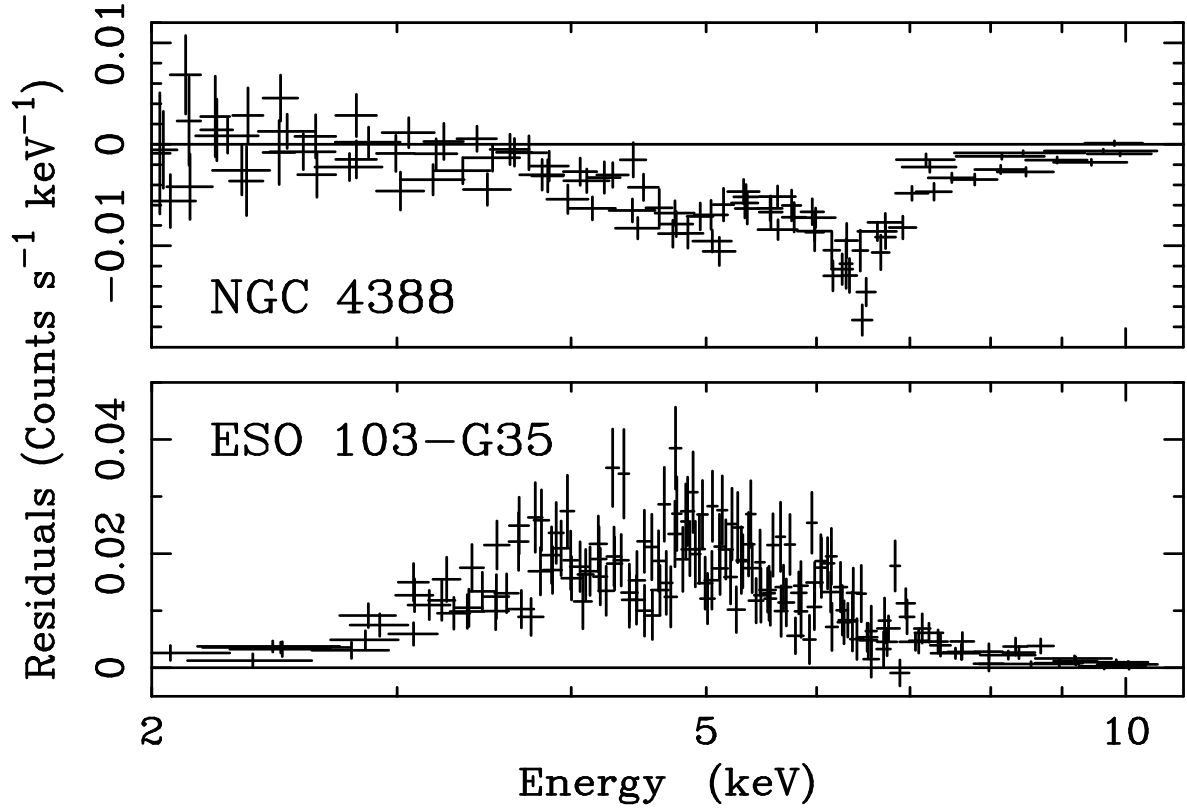
\noindent Fig. 10 --- Hard X-ray variability of NGC~4388 and
ESO~103--G35, both GIS and SIS data are shown. {\it Top} -- The
residuals from applying the 1993 model (Table 2a) to the 1995 data of
NGC~4388. The variability in the strength of the Fe K$\alpha$ emission
line near 6.4 keV can clearly be seen. {\it Bottom} -- As above but
applying the 1994 model (Table 2b) to the 1996 data of ESO~103--G35.

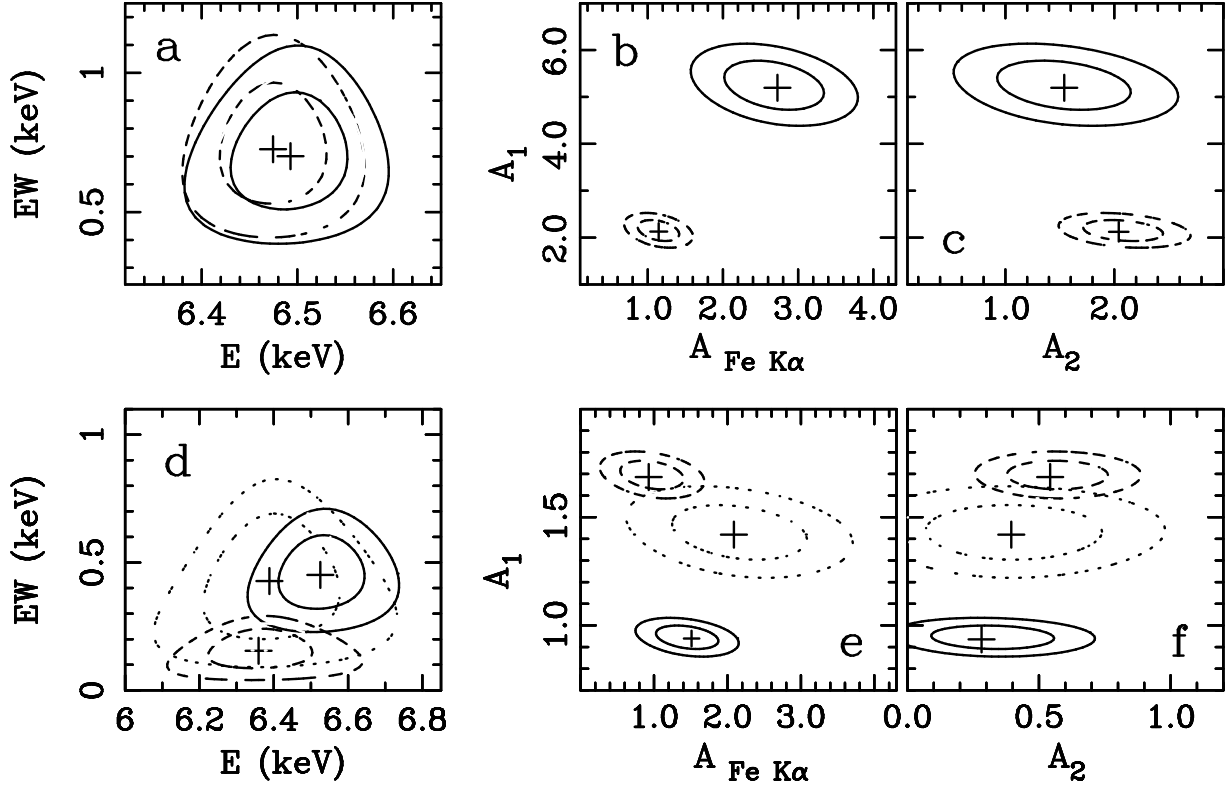
\noindent Fig. 11 --- The confidence contours of the fixed parameter
model in Tables 2a,b. The contours are the 68\% and 99\% confidence
contours for 4 interesting parameters. {\it Top panels} --- NGC~4388,
{\it solid line} = 1993 spectrum, {\it dashed line} = 1995 spectrum.
{\it Bottom panels} --- ESO~103--G35, {\it solid line} = 1994
spectrum, {\it dotted line} = 1995 spectrum, {\it dashed line} = 1996
spectrum.  (a,d) Emission line peak energy (keV) vs. emission line
equivalent width (keV). (b,e) Total photons held in the emission line
($10^{-4}$ photons cm$^{-2}$ s$^{-1}$) vs. normalization of the
absorbed power law continuum ($10^{-3}$ (b) $10^{-2}$ (e) 
photons cm$^{-2}$ s$^{-1}$
keV$^{-1}$ at 1 keV). (c,f) Normalization of the unabsorbed power law
continuum (A$_{2} \times 10^{-4}$ photons cm$^{-2}$ s$^{-1}$ keV$^{-1}$ at 1
keV) vs.  normalization of the absorbed power law continuum (A$_{1}
\times 10^{-3}$ photons cm$^{-2}$ s$^{-1}$ keV$^{-1}$ at 1 keV).
\clearpage

\end{document}